\documentclass[preprint,12pt]{elsarticle}

\usepackage{amssymb}
\usepackage[table]{xcolor}
\usepackage{amsmath}
\usepackage{url}

\journal{Elsevier}
\usepackage{booktabs}
\usepackage[table]{xcolor}
\usepackage{amssymb} 
\usepackage{amsmath}
\usepackage{subcaption}
\usepackage{tikz}
\usepackage[ruled,vlined]{algorithm2e}
\usepackage{pgfplots}
\usepackage{pgfplotstable}
\usepackage{float}
\usepackage{tikz}
\usetikzlibrary{calc}
\usepackage{amsmath, amssymb}
\usetikzlibrary{positioning, arrows.meta}
\usepackage{graphicx}
\usepackage{subcaption}

\usepackage{algpseudocode} 
\usepackage{pgfplots}
\pgfplotsset{compat=1.18}
\usetikzlibrary{patterns, decorations.pathmorphing}
\pgfplotsset{compat=1.18}
\usetikzlibrary{arrows.meta, positioning}
\usepackage{tikz}
\usepackage{graphicx}
\usepackage{xcolor} 
\usepackage{amsmath, amssymb} 
\usetikzlibrary{arrows.meta, positioning, shapes.geometric}
\begin{document}

\begin{frontmatter}




\title{Scaling of Interface Variability in Aerospace Composite Structures: A Multi-scale Stochastic Approach from Local Identification to Panel Response}
\author[AIRBUS,ONERADTIS]{Elton Donfack-Siewe} 
\author[ONERADTIS]{Sylvain Dubreuil}
\author[ONERADMAS]{Christian Fagiano}
\author[ONERADTIS]{Jérôme Morio}
\author[AIRBUS]{Jean-Philippe Navarro}
\affiliation[AIRBUS]{organization={AIRBUS OPERATIONS SAS},
            addressline={316 route de Bayonne}, 
            city={Toulouse},
            postcode={31060}, 
            country={France}}
\affiliation[ONERADTIS]{organization={ONERA/DTIS},
            addressline={Université de Toulouse}, 
            city={Toulouse},
            postcode={31055}, 
            country={France}}
\affiliation[ONERADMAS]{organization={ONERA/DMAS},
            addressline={Université Paris Saclay}, 
            city={Châtillon},
            postcode={92320}, 
            country={France}}
\begin{abstract}
Quantifying random spatial variations in composite interface properties is an essential prerequisite for the high-fidelity propagation of material properties variabilities through  structural numerical models. This study proposes a multi-scale stochastic framework designed to bridge the gap between local interfacial characterization at the coupon scale and the global response of industrial aerospace panels, focusing specifically on the skin-stringer bond line. Building upon previous stochastic modeling and identification methodologies for Mode I fracture, we first extend the likelihood-free Bayesian calibration to Mode II configurations to characterize interfacial stochastic parameters using experimental coupon data. To address the scale-up challenge, we employ a simulation-based Stochastic Finite Element Method (SFEM) to contrast two distinct probabilistic strategies—spatially uniform variations (Paradigm 2) and spatially correlated random fields (Paradigm 3)—against the traditional industrial B-basis standard (Paradigm 1). The numerical results reveal a dual mechanism in uncertainty propagation depending on the targeted airframe design criterion. For global performance, such as the ultimate load capacity, a significant scale effect leads to an uncertainty attenuation phenomenon; statistical averaging smooths out local heterogeneities, demonstrating that a computationally efficient spatially uniform approach is sufficient. Conversely, for local integrity, such as the threshold and spatial localization of early damage initiation, the inclusion of spatially correlated random fields becomes indispensable to capture premature degradation driven by local material weak spots. This balanced probabilistic treatment offers a route toward test derisking and predictive uncertainty management in the design of damage-tolerant aerospace composite airframes.

\end{abstract}

\begin{graphicalabstract}
\end{graphicalabstract}

\begin{highlights}
\item Quantification of manufacturing-induced interface variability using spatially correlated random fields.
\item Multiscale stochastic framework links mesoscale properties to panel damage initiation, propagation, and final collapse.

\item Scale-up effect leads to global stability despite high local material scatter.

\item Evaluation of alternative probabilistic strategies against the traditional industrial B-basis standard.

\end{highlights}

\begin{keyword}
Multi-scale analysis \sep Random Fields \sep  Stiffened Panel \sep Bayesian identification \sep Composite Interface


\end{keyword}

\end{frontmatter}



\section{Introduction}

\label{sec:introduction}

Composite materials are fundamental to modern aerospace structures due to their exceptional specific mechanical properties~\cite{Yu2016}. However, their deployment is complicated by intrinsic heterogeneity and anisotropy~\cite{cinquin2002}, stemming from their multiscale architecture~\cite{Laboulfie2023}. Manufacturing processes—such as vacuum molding or autoclaving—introduce variability and random spatial variation in mechanical properties ~\cite{Grimsley2001}. Structural analysis using the traditional Finite Element Method (FEM) in this case cannot account for all possible
output scenarios. While traditional design methodologies mitigate these uncertainties through safety factors~\cite{CMH17}, the industry investigates the use of physics-based uncertainty quantification.

This multi-scale uncertainty propagation challenge directly maps onto the traditional aerospace Certification Pyramid (also known as the Building Block Approach) dictated by airworthiness regulations such as EASA CS-25 \cite{easa_cs25} and FAA FAR 25 \cite{faa_far25} for primary structures. Moving from experimental characterization at the coupon scale (the pyramid base) to complex components like stiffened panels (the structural element level) traditionally involves substantial financial and time investments.

A major difficulty in aerospace design is the assembly of composite parts, especially the interface between the fuselage skin and the stringer. In stiffened panels, the overall structural integrity is often dictated more by the behavior of the adhesive or co‑cured bond line than by the bulk material itself~\cite{Yu2016, Meeks2005}. Therefore, the reliability of the entire assembly hinges on accurately predicting the initiation and propagation of disbonding at this specific interface. Present practices usually represent this connection with Cohesive Zone Models (CZM) that assume spatially uniform material parameters—fracture energy $G_c$ and peak strength $\sigma_{max}$~\cite{AbaqusCohesive, Alfano2006}.

Nevertheless, modelling the skin‑stringer joint as a perfectly homogeneous interface contradicts actual produced part conditions. Standardised characterisation tests used in the industry—especially the Mode I and Mode II procedures—regularly show pronounced spatial scatter in fracture propagation and intricate R‑curve behaviours that constant‑parameter models fail to capture ~\cite{Pereira2004_ModeI,Pereira2003_ModeII,Kozik2017}.

To numerically account for these uncertainties, we recently proposed a stochastic framework at the coupon scale~\cite{DonfackSiewe2026_2}, replacing the assumption of a homogeneous interface with spatially correlated random fields~\cite{Vanmarcke1988} using the traditional Finite Element Method (FEM). This work demonstrated that the correlation length ($l_c$) is one of the primary drivers of structural response variability at the coupon level, while the mathematical form of the covariance kernel remains a secondary factor. The identification of this statistical spatial parameter, along with the constitutive cohesive zone properties, was achieved via an Approximate Bayesian Computation (ABC-kNN) approach~\cite{Thomas2012}. By applying this methodology to standard Mode I fracture tests—specifically the Double Cantilever Beam (DCB) configuration—the framework captured the experimental dispersion. Building on this foundation, the present study first extends this identification methodology to the Mode II configuration, allowing for a complete characterization of the interfacial stochastic parameters before transitioning to larger structural scales.

However, one fundamental question remains regarding the scale effect (transition from coupon to panel). It is currently unclear whether the local material interface heterogeneities identified at the coupon level are smoothed out or, conversely, amplified when projected onto large-scale structural components. Addressing this "scale-up" challenge is important for the predictive management of uncertainties in the design of damage-tolerant composite airframes.

To address this challenge, this study presents an integrated stochastic numerical framework evaluated on a representative industrial stiffened panel. Rather than treating a single method in isolation, we contrast two distinct stochastic representation strategies calibrated from experimental coupon data: a spatially uniform (constant) approach accounting for global specimen-to-specimen variations, and a spatially correlated random fields approach capturing local intra-bondline heterogeneities. Both advanced probabilistic formulations are systematically evaluated against the classical industrial standard, which relies on a conservative approach using deterministic B-basis values determined via Wilks' formula~\cite{Wilks1941} obtained by statistical analysis from coupon test data. By employing a simulation-based Stochastic Finite Element Method (SFEM), we project these alternative descriptions onto the full scale mesh of the structure of interest to evaluate how different uncertainty treatments govern skin buckling, disbondinginitiation, and final structural collapse. This statistical approach aligns with recent trends in computational mechanics~\cite{Vinot2023, Stefanou2019, VanBavel2023}. Notably, recent studies have demonstrated how random microstructure morphology significantly influences the macroscopic bending behavior of composite plates, further justifying the need for stochastic frameworks~\cite{Gavallas2025}.

The paper is organized as follows. Section \ref{section_2} establishes the computational case study and formulates the three uncertainty paradigms, while Section \ref{section_3} details the Bayesian identification strategy developed at the coupon scale. Section \ref{section_4} investigates the scale-up propagation and structural response variability at the macroscale. Finally, Section 5 provides conclusions and future perspectives.

\section{Computational case study and formulation of three uncertainty paradigms}
\label{section_2}
To investigate the propagation of interface uncertainties across scales, this section establishes a multi-scale framework evaluated on a representative industrial aerospace structure. Rather than treating a single stochastic method in isolation, the problem is formulated to contrast three distinct uncertainty management paradigms.

\subsection{Industrially representative stiffened panel configuration}
\begin{figure}[H]
    \centering
    \includegraphics[width=1.0\textwidth]{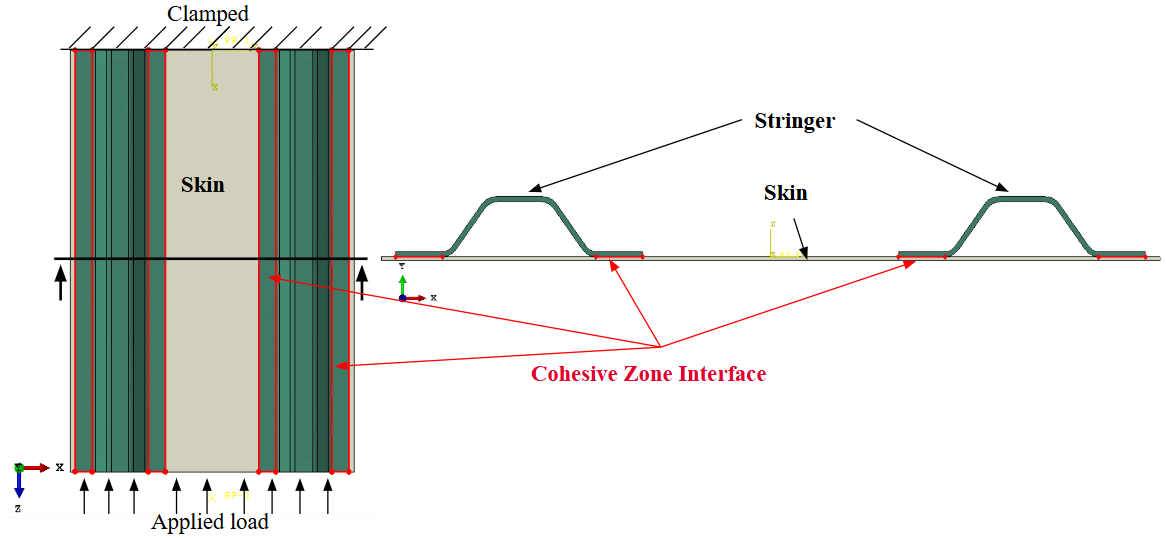}
    \caption{Stiffened panel configuration under compression: Boundary conditions and Cohesive Zone Modeling (CZM) locations.}
    \label{fig:panel_setup}
\end{figure}

The structural case study consists of an aerospace-grade stiffened panel comprising a base skin co-cured or bonded to two omega-shaped stringers illustrated in Figure~\ref{fig:panel_setup}. To ensure a representative evaluation of scale-up effects, the panel involves a substantial dimensional increase compared to standard coupon specimens. The length and width ratios between the structural scales are defined as:

\begin{equation}
\frac{L_{panel}}{L_{coupon}} \approx 3, \quad \frac{W_{panel}}{W_{coupon}} \approx 15,
\end{equation}
where the longitudinal coordinate $L$ is aligned with the compressive loading axis, and $W$ denotes the total transverse width of the skin-stringer assembly.

The high-fidelity non-linear finite element model is implemented using the Abaqus implicit solver. The composite laminates are discretized using 3D continuum shell elements, while the skin-stringer bond lines are modeled via $\text{COH3D8}$ zero-thickness cohesive elements. 

To preserve numerical consistency and eliminate artificial stiffness gradients, a double-conforming node-to-node strategy is deployed. This approach acts as a topological bridge ensuring that the lower-face nodes are perfectly coincident with the skin mesh and the upper-face nodes match the stringer mesh configuration. Consequently, this conforming discretization preserves the exact mesh density and element formulations (shell and cohesive) calibrated at the coupon scale, maintaining a constant characteristic mesh size across scales.

The local interfacial degradation is represented by a bilinear traction-separation law governed by the critical fracture energy vector $\mathbf{G_c} = [G_{Ic}, G_{IIc}]^T$, where $G_{Ic}$ represents the critical energy release rate for the Mode I opening configuration and $G_{IIc}$ represents the critical energy release rate for the Mode II sliding shear configuration. The interfacial peak strength vector is defined as $\boldsymbol{\sigma_c} = [\sigma_{n,c}, \sigma_{t,c}]^T$, where $\sigma_{n,c}$ denotes the critical cohesive strength in the normal direction and $\sigma_{t,c}$ denotes the critical cohesive strength in the tangential shear direction. Damage initiation is determined by a quadratic stress criterion based on these cohesive joint strengths, while the subsequent mixed-mode progressive damage softening is resolved via the Benzeggagh–Kenane (BK) criterion \cite{Benzeggagh1996}, where the material mixity parameter is fixed to $\eta = 2$ \cite{AbdelMonsef2023}.

The panel is subjected to a progressive compressive displacement extending well into the post-buckling regime until ultimate structural collapse. While global manufacturing uncertainties (such as geometric tolerances and ply mismatches) are fixed in this study, the analysis isolates how interfacial fracture variations govern the structural failure kinematics. 

\begin{figure}[H]
     \centering
     \begin{subfigure}[b]{0.48\textwidth}
         \centering
         \includegraphics[width=\textwidth]{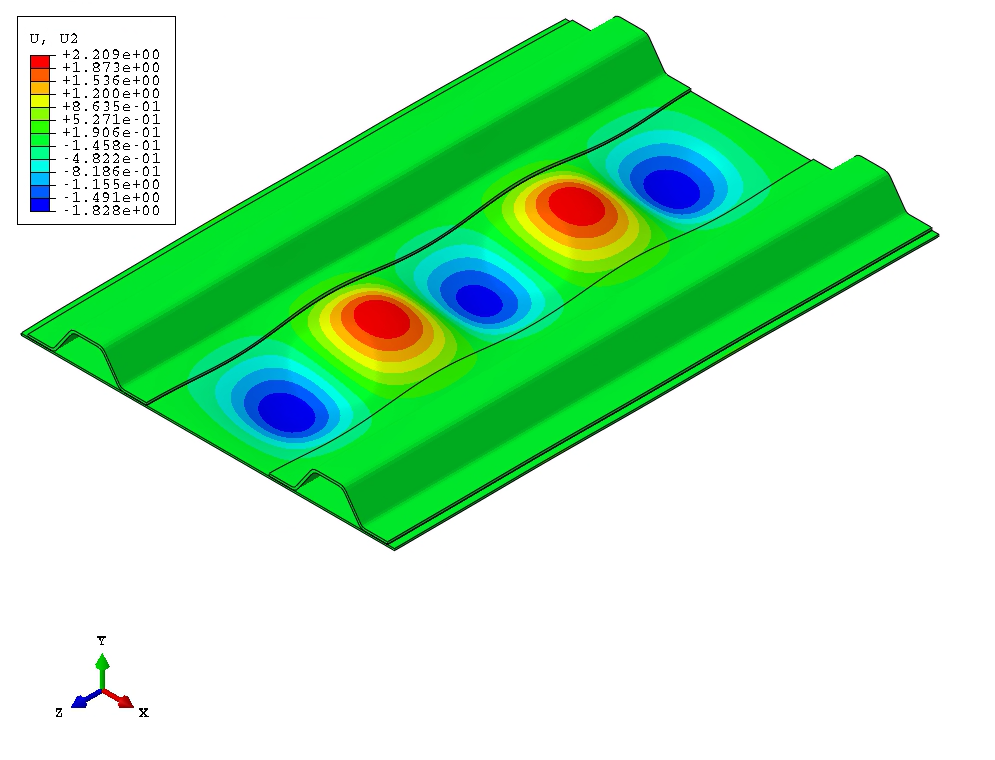}
         \caption{Isometric view of skin buckling between stringers.}
         \label{fig:skin_1}
     \end{subfigure}
     \hfill
     \begin{subfigure}[b]{0.48\textwidth}
         \centering
         \includegraphics[width=\textwidth]{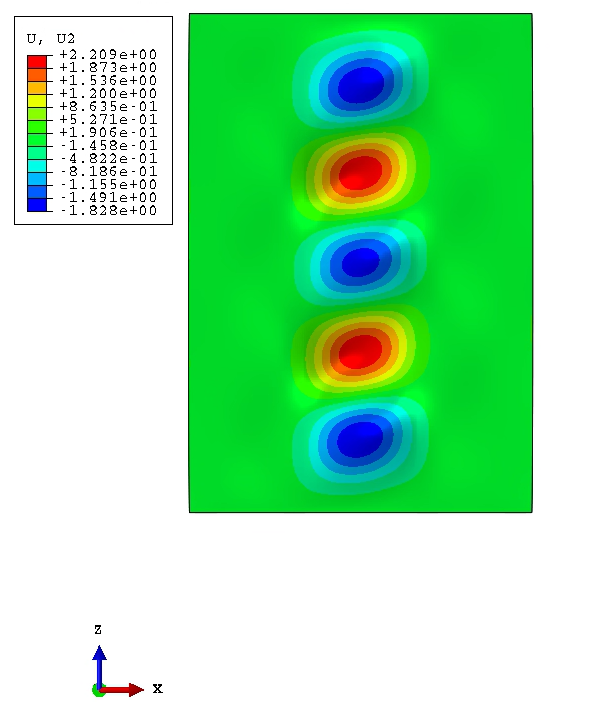}
         \caption{Skin buckling profile in the (X,Z) plane.}
         \label{fig:skin_2}
     \end{subfigure}
     \caption{Representative skin buckling mode between the two stringers.}
     \label{fig:skin}
\end{figure}

\begin{figure}[H]
     \centering
     \begin{subfigure}[b]{0.48\textwidth}
         \centering
         \includegraphics[width=\textwidth]{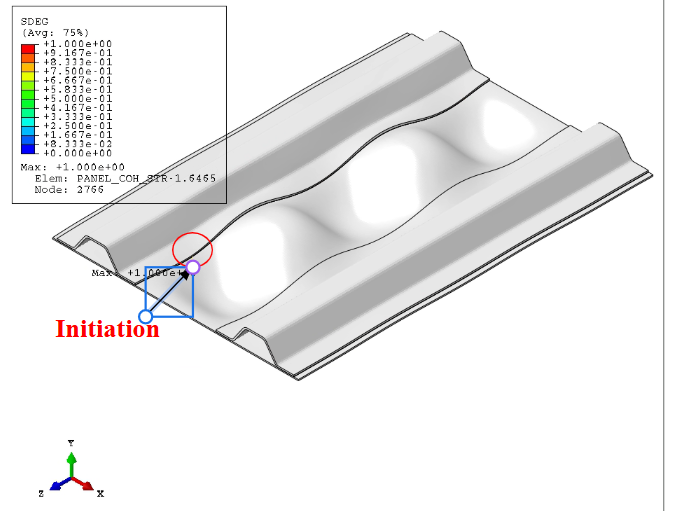}
         \caption{Global view of initial disbondinglocalization.}
         \label{fig:SDEG_1}
     \end{subfigure}
     \hfill
     \begin{subfigure}[b]{0.48\textwidth}
         \centering
         \includegraphics[width=\textwidth]{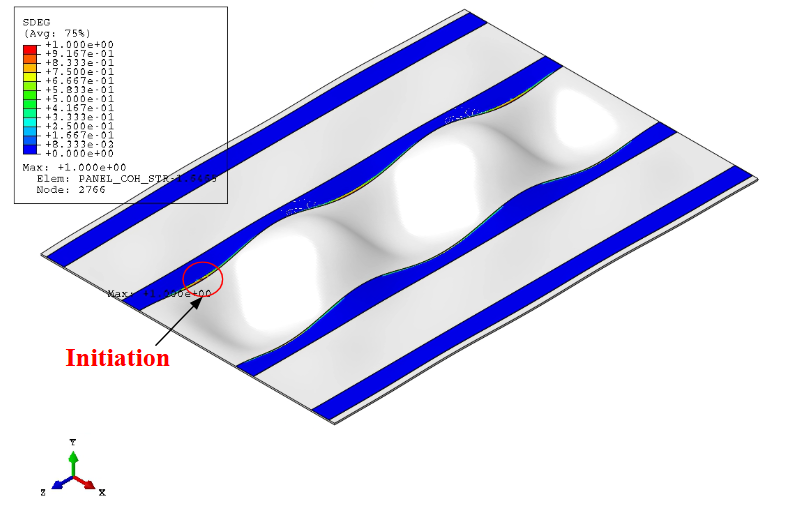}
         \caption{Detailed view with stringers removed for clarity.}
         \label{fig:SDEG_2}
     \end{subfigure}
    \caption{Localization of the first disbondingevent ($d_{max} = 1$).}
     \label{fig:SDEG}
\end{figure}

\begin{figure}[H]
     \centering
     \begin{subfigure}[b]{0.48\textwidth}
         \centering
         \includegraphics[width=\textwidth]{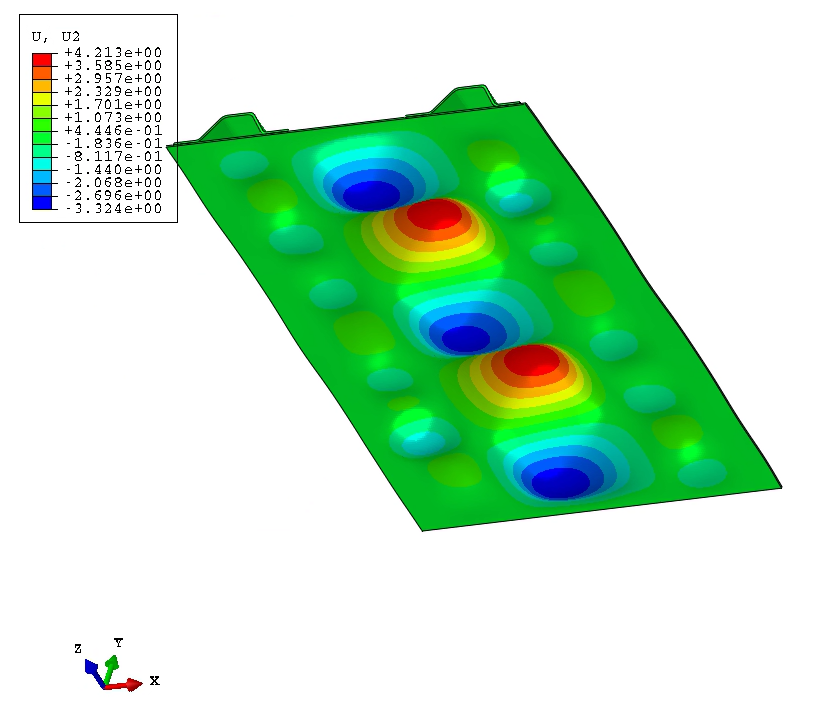}
         \caption{Isometric view of skin buckling under the stringer.}
         \label{fig:str_1}
     \end{subfigure}
     \hfill
     \begin{subfigure}[b]{0.48\textwidth}
         \centering
         \includegraphics[width=\textwidth]{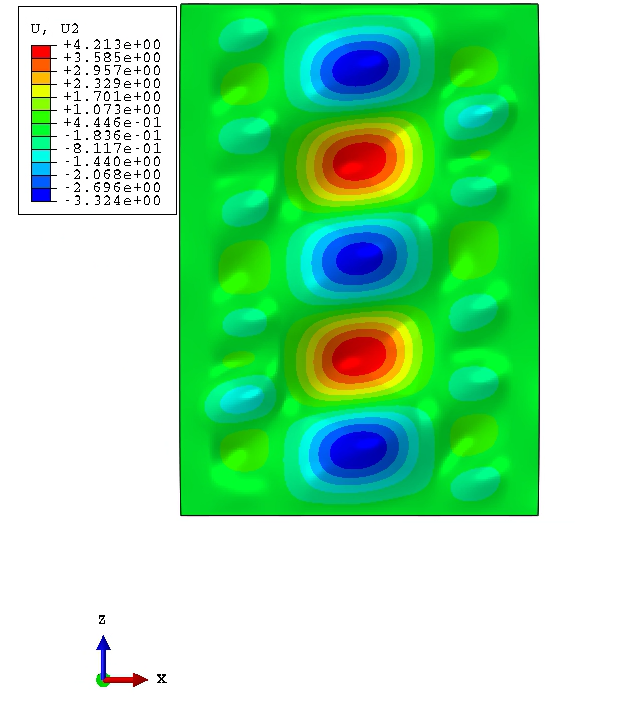}
         \caption{Buckling profile in the (X,Z) plane.}
         \label{fig:str_2}
     \end{subfigure}
     \caption{Representative skin buckling mode occurring under the omega stringers.}
     \label{fig:str_buckling}
\end{figure}

\begin{figure}[H]
     \centering
     \begin{subfigure}[b]{0.48\textwidth}
         \centering
         \includegraphics[width=\textwidth]{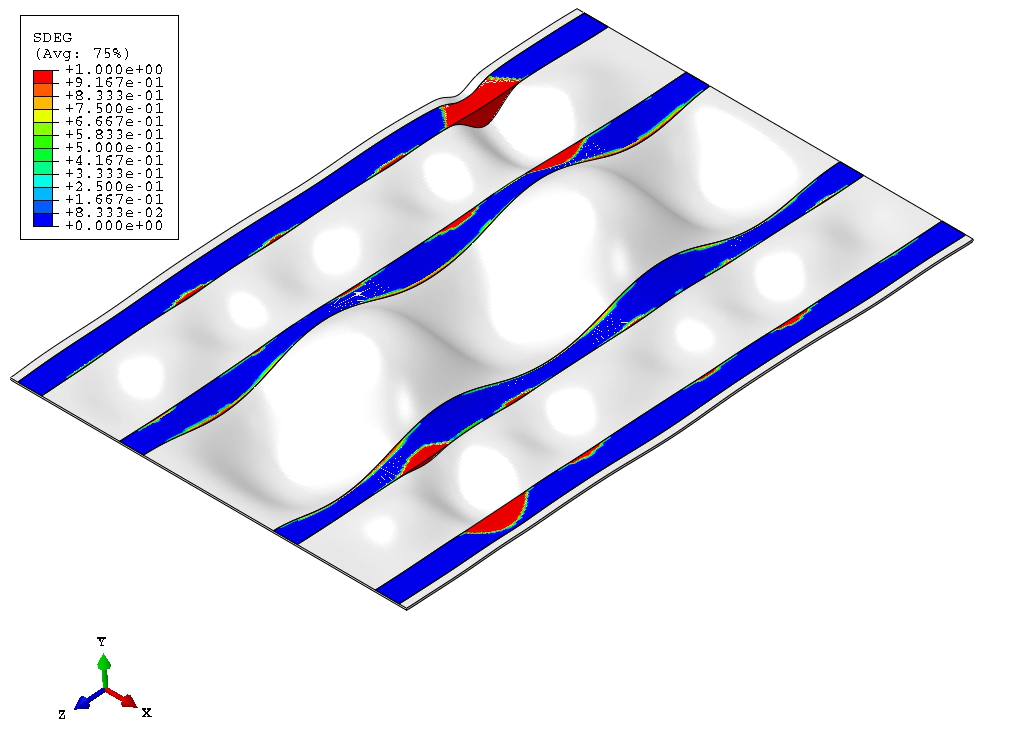}
         \caption{disbondingarea at peak load ($F_{max}$).}
         \label{fig:debonding_max}
     \end{subfigure}
     \hfill
     \begin{subfigure}[b]{0.48\textwidth}
         \centering
         \includegraphics[width=\textwidth]{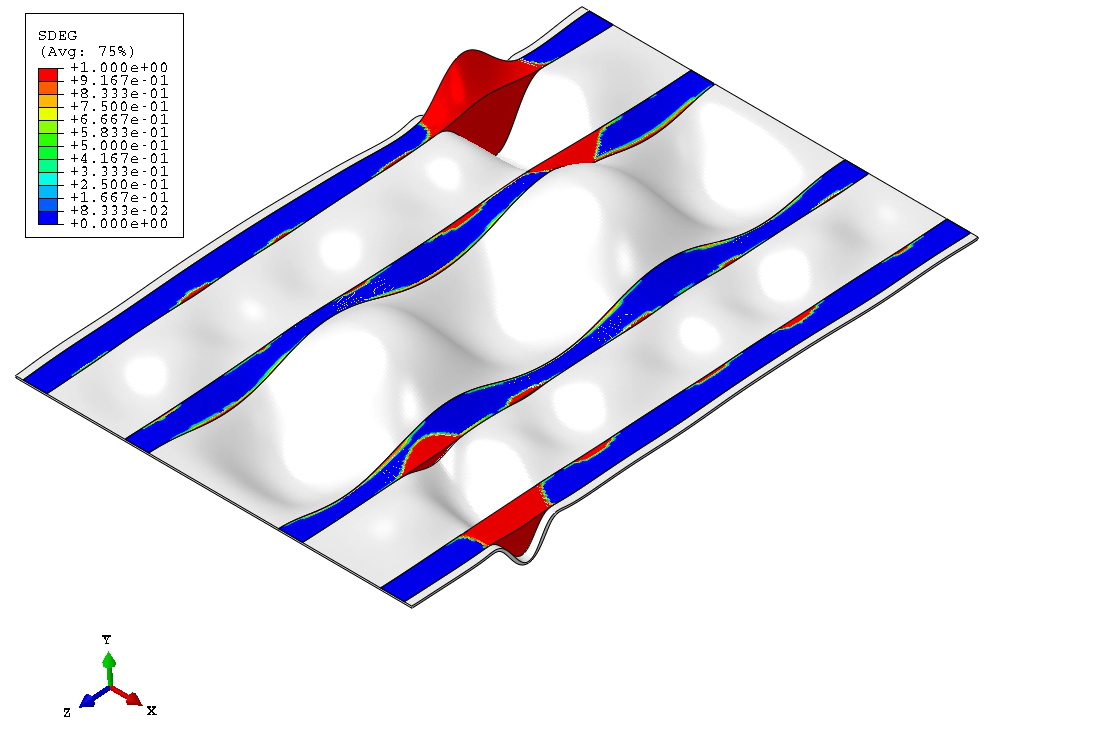}
         \caption{disbondingarea in the post-peak regime.}
         \label{fig:debonding_after}
     \end{subfigure}
     \caption{Numerical simulation of the disbondingarea: (a) state at maximum force $F_{max}$ and (b) post-peak state at $1.01 \times u_c$, where $u_c$ is the displacement corresponding to $F_{max}$.}
     \label{fig:global_debonding}
\end{figure}

The primary output performance indicators evaluated comprise:
\begin{itemize}
    \item The skin buckling force ($F_{buckling}$), which marks the load corresponding to the initial local out-of-plane instability of the skin section located between the two stringers as illustrated in Figure~\ref{fig:skin}.
    \item The initial disbondingforce ($F_{init}$), defined as the compressive load where the localized interfacial damage variable first reaches unity ($d_{max} = 1$) at any point across the four cohesive joint lines as exemplified in Figure~\ref{fig:SDEG}.
    \item The skin buckling force under the stringers ($F_{str\_buckling}$), which marks a critical transition in the post-buckling regime where the skin directly beneath the omega stringers begins to buckle (example in Figure~\ref{fig:str_buckling}), generating highly critical interface peeling stresses.
    \item The global peak force ($F_{max}$), which defines the ultimate load-bearing capacity of the panel prior to catastrophic collapse and total skin-stringer decohesion, an event captured in Figure~\ref{fig:global_debonding}.
\end{itemize}

\subsection{Mathematical formulation of the three uncertainty paradigms for interface properties}
Let $\mathbf{\theta}(\mathbf{x}) = [\mathbf{G_c}(\mathbf{x})^T, \boldsymbol{\sigma_c}(\mathbf{x})^T]^T = [G_{Ic}(\mathbf{x}), G_{IIc}(\mathbf{x}), \sigma_{n,c}(\mathbf{x}), \sigma_{t,c}(\mathbf{x})]^T$ denote the joint vector of local cohesive zone parameters evaluated at the two-dimensional spatial coordinate vector $\mathbf{x} \in \mathbb{R}^2$ along the skin-stringer interface plane. Here, $\mathbf{G_c}(\mathbf{x})$ represents the space-dependent critical fracture energy vector containing the Mode I ($G_{Ic}$) and Mode II ($G_{IIc}$) components, while $\boldsymbol{\sigma_c}(\mathbf{x})$ represents the interfacial peak strength vector containing the normal ($\sigma_{n,c}$) and tangential ($\sigma_{t,c}$) components. To evaluate the conservatism and predictive capability of alternative uncertainty treatments, three distinct paradigms are formulated:

\subsubsection{Paradigm 1: B-Basis deterministic uniform propagation (industrial regulatory standard)}
Following standard aerospace certification guidelines (CMH-17~\cite{CMH17}), this approach reduces material variability to a deterministic lower-bound design value. The interface properties are assumed perfectly homogeneous across the entire spatial domain and are set to their non-parametric B-basis values:
\begin{equation}
\mathbf{\theta}(\mathbf{x}) = \mathbf{\theta}_{B},
\end{equation}
where $\mathbf{\theta}_{B}$ is an invariant vector determined using Wilks' formula \cite{Wilks1941} based on a target population survival probability extracted from the experimental coupon test data.

\subsubsection{Paradigm 2: spatially uniform properties (stochastic baseline case)}
This formulation introduces probabilistic uncertainty by treating the parameter vector $\mathbf{\Theta}_{c} = [G_{Ic}, G_{IIc}, \sigma_{n,c}, \sigma_{t,c}]^T$ as a global random variable that fluctuates from panel to panel, while remaining strictly constant and uniform across the entire interface for any given structural realization. The interface properties are space-invariant and governed by a joint probability density function $f_{\Theta}$:

\begin{equation}
\mathbf{\theta}(\mathbf{x}) = \mathbf{\Theta}_{c} \sim f_{\Theta}(\cdot).
\end{equation}
This baseline case effectively captures specimen-to-specimen macrostructural variability while completely neglecting local intra-bondline heterogeneities. The generation of these random samples is performed using a classical Monte Carlo approach.

\subsubsection{Paradigm 3: spatially correlated Random Fields}
In distinct contrast to the space-invariant formulation of Paradigm 2 ($\mathbf{\theta}(\mathbf{x}) = \mathbf{\Theta}_c$), this approach explicitly accounts for intra-bondline heterogeneities by modeling the cohesive properties as a continuous spatial random field indexed by the coordinate vector $\mathbf{x} \in \mathbb{R}^2$. Following the framework in Donfack-Siewe et al. \cite{DonfackSiewe2026_2}, the property vector $\mathbf{\theta}(\mathbf{x})$ is treated as a stationary and ergodic stochastic process characterized by a normal marginal distribution and a spatial dependence structure governed by the correlation length vector $\mathbf{l}_c = [l_c^I, l_c^{II}]^T$ \cite{Vanmarcke1988} which defines how quickly the material properties change across the joint for both Mode I (opening) and Mode II (shearing).

To implement this framework within a simulation-based Stochastic Finite Element Method (SFEM), each individual cohesive property $\theta_k$ (for $k \in \{1, 2, 3, 4\}$, two energies and two stresses) undergoes spatial discretization independently. The domain is evaluated at $m$ mesh elements $\mathbf{x}_1, \dots, \mathbf{x}_m \in \mathbb{R}^2$. Gathering the localized values for a given property yields a random vector $\mathbf{\Theta}_k = [\theta_k(\mathbf{x}_1), \dots, \theta_k(\mathbf{x}_m)]^T \in \mathbb{R}^{m}$, whose multivariate normal joint probability density function $f(\mathbf{\Theta}_k)$ is expressed as:

\begin{equation}
    f(\mathbf{\Theta}_k) = \frac{1}{(2\pi)^{m/2} \det(\mathbf{\Sigma}_k)^{1/2}} \exp\left( -\frac{1}{2} (\mathbf{\Theta}_k - \mu_{k}\mathbf{1}_m)^T \mathbf{\Sigma}_k^{-1} (\mathbf{\Theta}_k - \mu_{k}\mathbf{1}_m) \right),
\end{equation}where $\mu_{k}$ is the constant baseline mean value for the $k$-th property, $\mathbf{1}_m \in \mathbb{R}^m$ is a vector of ones, and $\mathbf{\Sigma}_k \in \mathbb{R}^{m \times m}$ is the spatial covariance matrix. The coefficients of $\mathbf{\Sigma}_k$ are populated via an analytical spatial exponential covariance kernel evaluating the Euclidean distance between any two discrete mesh coordinate locations $\mathbf{x}_i$ and $\mathbf{x}_j$:\begin{equation}\Sigma_{k, ij} = C_k(\mathbf{x}_i, \mathbf{x}_j ; l_{c,k}) = S_{k}^2 \exp \left( - \frac{|\mathbf{x}_i - \mathbf{x}_j|}{l_{c,k}} \right).\end{equation}
The stochastic field for each property is fully defined by a triad of scalar hyperparameters: the mean value $\mu_{k}$, the standard deviation $S_{k}$ (scatter magnitude), and the correlation length $l_{c,k}$ (spatial decay scale).

To generate explicit realizations of this discretized random vector within a classical Monte Carlo Simulation (MCS) framework, a standard approach relies on the Cholesky decomposition \cite{Cholesky1924} of the spatial covariance matrix, defined as $\mathbf{\Sigma}_k = \mathbf{L}_k \mathbf{L}_k^T$, where $\mathbf{L}_k \in \mathbb{R}^{m \times m}$ is a lower triangular matrix. Individual sample realizations are then computed directly using the following linear transformation:
\begin{equation}
    \mathbf{\Theta}_k = \mu_{k}\mathbf{1}_m + \mathbf{L}_k \mathbf{\xi},
\end{equation}
where $\mathbf{\xi} \in \mathbb{R}^m$ denotes a random vector composed of independent standard normal random variables. While straightforward to implement, this direct factorization scales with an $\mathcal{O}(m^3)$ computational complexity. Consequently, when the mesh refinement parameter $m$ becomes very large, MCS encounters severe numerical bottlenecks and prohibitive memory limitations. To circumvent this curse of dimensionality and ensure numerical tractability during large-scale simulations, recourse to the Karhunen-Loève (KL) \cite{Ghanem1991} expansion is preferred. The KL framework expands the continuous random field via a truncated series of orthogonal deterministic eigenfunctions and eigenvalues, drastically compressing the stochastic space by retaining only the highest-energy modes.

\section{Mesoscale parameter identification: Bayesian calibration of interfacial stochastic models}
\label{section_3}

This section details the uncertainty identification strategy designed to bridge experimental coupon-scale observations with global macrostructural assessment. The primary objective is to calibrate the parameters governing interfacial degradation and map them systematically onto the three alternative uncertainty paradigms established in Section~\ref{section_2}.

\subsection{General Bayesian inverse framework at the coupon scale}

The identification of interface properties relies on a statistical inverse framework designed to update parameter knowledge based on experimental observations. Let $\boldsymbol{\beta} \in \mathbb{R}^p$ denote the vector of hyperparameters to be identified. Following Bayes' theorem, the posterior probability density function $\pi(\boldsymbol{\beta} \mid Z_{\mathrm{obs}})$ given an experimental observation $Z_{\mathrm{obs}}$ is formulated as:

\begin{equation}
    \pi(\boldsymbol{\beta} \mid Z_{\mathrm{obs}}) = \frac{\pi(Z_{\mathrm{obs}} \mid \boldsymbol{\beta})\, \pi(\boldsymbol{\beta})}{\pi(Z_{\mathrm{obs}})} \propto \pi(Z_{\mathrm{obs}} \mid \boldsymbol{\beta})\,\pi(\boldsymbol{\beta}),
    \label{eq:bayesRule}
\end{equation}
where $\pi(\boldsymbol{\beta})$ represents the prior distribution constructed from physical constraints or expert knowledge, and $\pi(Z_{\mathrm{obs}} \mid \boldsymbol{\beta})$ is the likelihood function. 

In experimental mechanics, characterization is rarely limited to a single test. To capture specimen-to-specimen and intra-specimen variabilities, a dataset containing multiple independent coupons is required. Let $\widetilde{Z}_{\mathrm{obs}} = \{ Z_{\mathrm{obs}}^{(j)}\}_{j=1}^{N_{\mathrm{test}}}$ denote the set of available experimental data. To incorporate information from the entire campaign, the global posterior is obtained by aggregating the individual posterior approximations~\cite{DonfackSiewe2026_2}:
\begin{equation}
    \pi(\boldsymbol{\beta} \mid \widetilde{Z}_{\mathrm{obs}}) = \frac{1}{N_{\mathrm{test}}} \sum_{j=1}^{N_{\mathrm{test}}} \pi(\boldsymbol{\beta} \mid Z_{\mathrm{obs}}^{(j)}).
    \label{eq:multi_posterior}
\end{equation}

\begin{equation}
    \pi(\boldsymbol{\Theta} \mid \widetilde{Z}_{\mathrm{obs}}) = \frac{1}{N_{\mathrm{test}}} \sum_{j=1}^{N_{\mathrm{test}}} \pi(\boldsymbol{\Theta} \mid Z_{\mathrm{obs}}^{(j)}).
    \label{eq:multi_posteriodzdr}
\end{equation}
This aggregation procedure naturally accounts for the scatter inherent in the experimental process while enforcing a consensus on the identified parameters.

\subsection{Bayesian calibration results}
The experimental dataset leverages standardized Mode I (Double Cantilever Beam - DCB) and Mode II (End-Notched Flexure - ENF) fracture tests, comprising $N_{\mathrm{test}}=5$ independent specimens per configuration (detailed in \ref{app:experiments}). Depending on the core focus of each paradigm, the Bayesian inverse problem is formulated and resolved as follows:

\begin{enumerate}
    \item \textbf{Paradigm 2 (spatially uniform random variables) via classical MCMC:} 
    In this baseline case, the parameter vector to identify is directly the space-invariant cohesive property vector itself, denoted as $\boldsymbol{\beta}= \mathbf{\Theta}_c$. Because the parameters are assumed uniform across the entire interface, the likelihood function remains computationally manageable. A classical Markov Chain Monte Carlo (MCMC) algorithm is deployed to sample and identify the joint probability density function $f_{\Theta}(\mathbf{\Theta}_c)$ based on the experimental curves.
    
    \item \textbf{Paradigm 1 (deterministic B-Basis) derived from MCMC:} 
    This regulatory framework requires a single lower-bound design value $\mathbf{\theta}_{B}$. Rather than calibrating this value independently, it is directly deduced from the probability law $f_{\Theta}$ identified via the MCMC process in Paradigm 2. By applying Wilks' non-parametric formula to the MCMC population, we compute the invariant vector $\mathbf{\theta}_{B}$ that guarantees a $99\%$ survival probability with a $95\%$ confidence level, complying with aerospace regulations (CMH-17) \cite{CMH17}. The calibration results for these spatially uniform properties are presented in Figure~\ref{MCMC1}.

\begin{figure}[H]
    \centering
    \hspace*{-1.5cm}    
    \includegraphics[height=0.53\textheight]{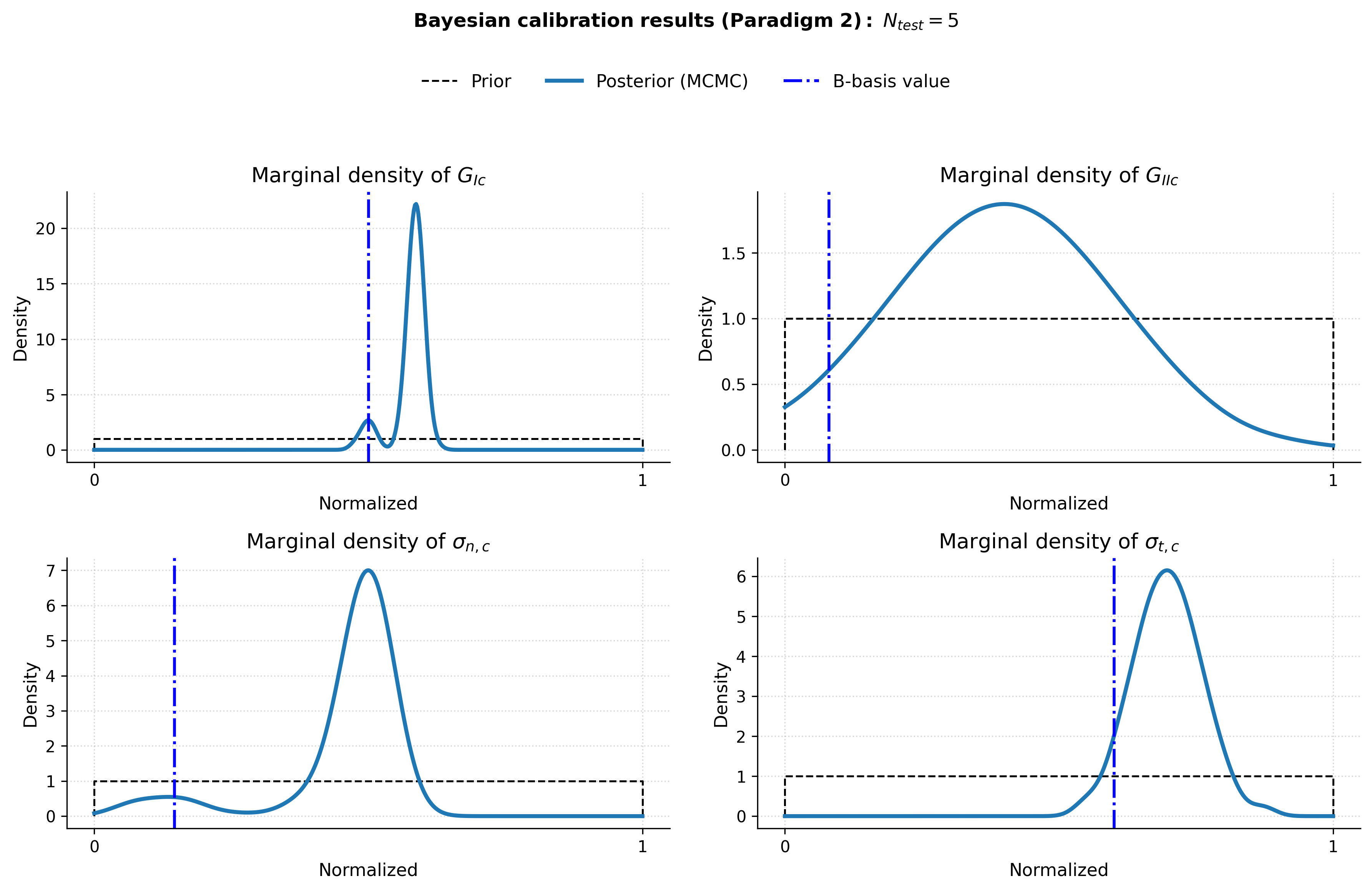}
    \caption{Posterior distributions of Mode I and Mode II spatially uniform interface parameters ($\mathbf{\Theta}_c$) calibrated via classical MCMC from $N_{\mathrm{test}}=5$ DCB tests and $N_{\mathrm{test}}=5$ ENF tests.}
    \label{MCMC1}
\end{figure}

Additionally, Figure~\ref{Val_MCMC} illustrates the validation of the calibrated uniform models through forward uncertainty propagation (comprising 100 Monte Carlo simulations) for both DCB and ENF test configurations. From these response curves, the capacity of the space-invariant paradigm to capture macrostructural specimen-to-specimen deviations can be verified.
    
\begin{figure}[H]
    \centering
    \begin{subfigure}[b]{0.48\textwidth}
    \centering
    \includegraphics[width=\textwidth]{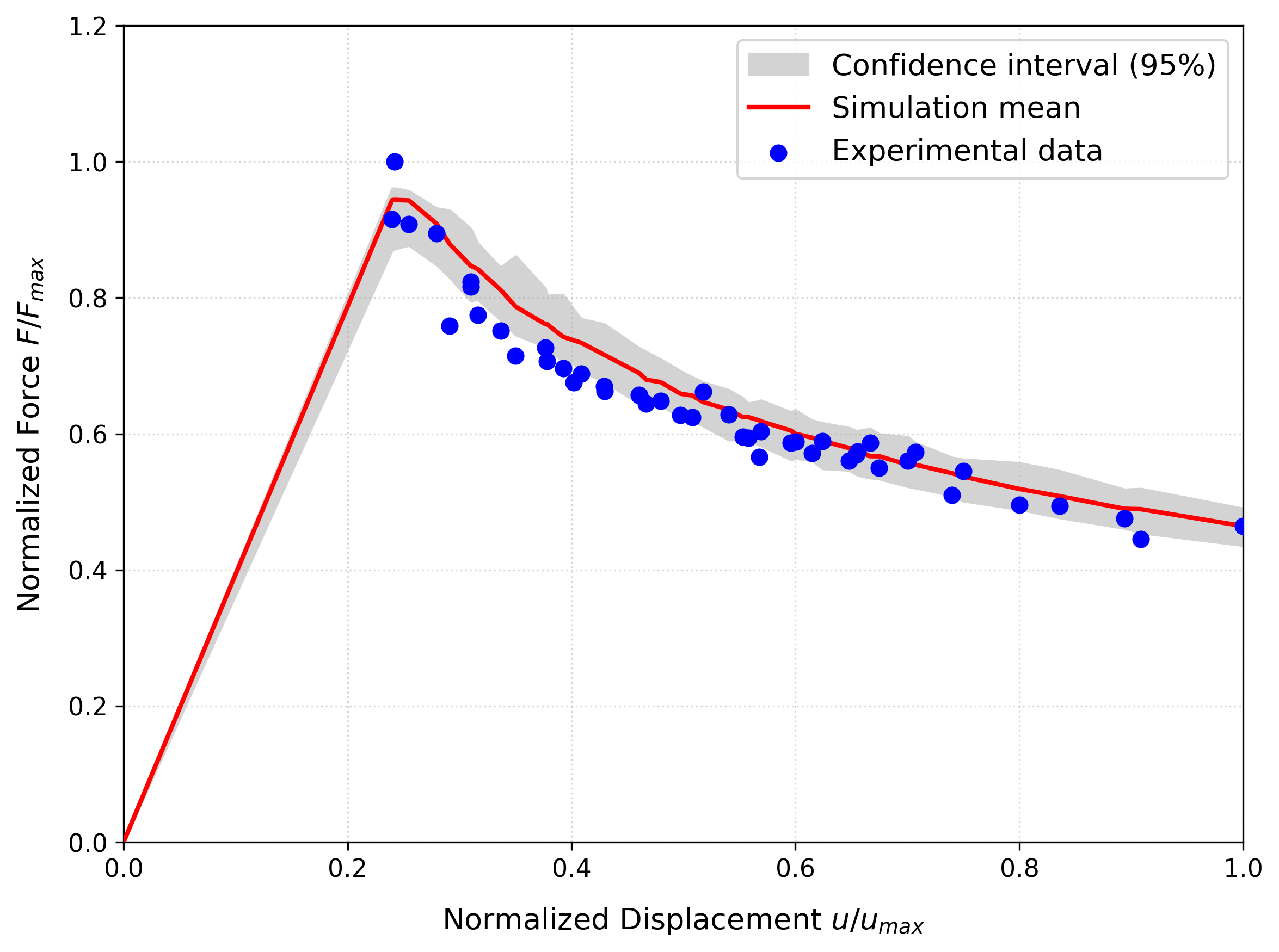}
    \caption{Mode I (DCB)}\label{fig:dcb_val}
    \end{subfigure}
    \hfill
    \begin{subfigure}[b]{0.48\textwidth}
    \centering
    \includegraphics[width=\textwidth]{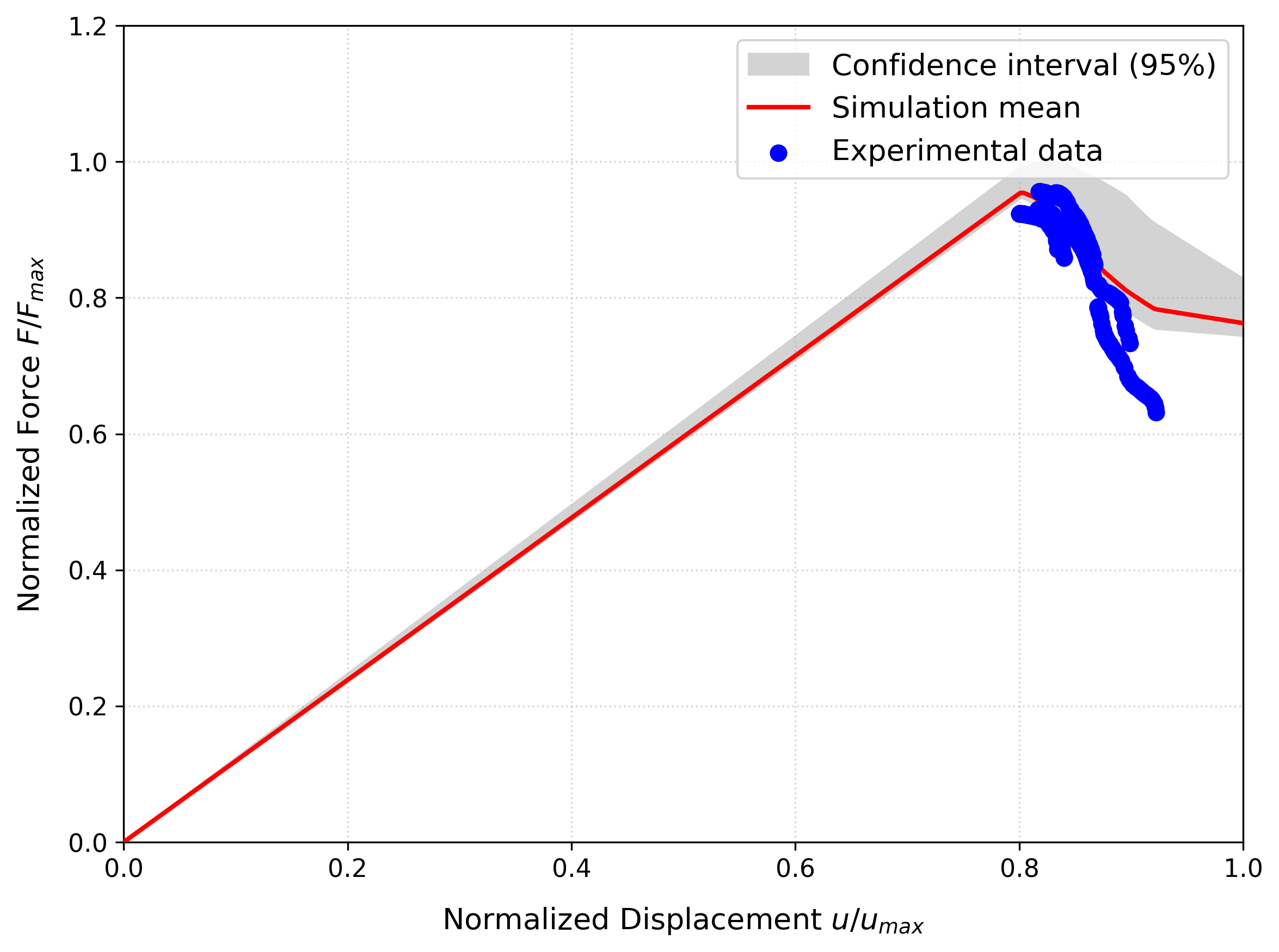}
    \caption{Mode II (ENF)}\label{fig:enf_val}
\end{subfigure}
\caption{Forward propagation of global macrostructural variability: 100 numerical realizations (grey lines) generated from the uniform posterior distribution, compared against experimental data (blue dots).}
\label{Val_MCMC}
\end{figure}
    
    \item \textbf{Paradigm 3 (spatially correlated random fields) via Approximate Bayesian Computation (ABC):} 
    In this approach, the parameters to identify are no longer the local property values, but the triad of random field hyperparameters gathered in the vector $\boldsymbol{\beta}_k = [\mu_k, S_k, l_{c,k}]^T$ (for $k \in \{1, 2, 3, 4\}$), where $\mu_k$ is the mean, $S_k$ is the standard deviation, and $l_{c,k}$ is the finite spatial correlation length.
    
    This configuration introduces extreme computational complexity. Because the continuous random field varies from point to point, evaluating an explicit likelihood function through non-linear finite element simulations is computationally prohibitive \cite{DonfackSiewe2026_2}. To circumvent this challenge, we leverage the likelihood-free Approximate Bayesian Computation (ABC) framework proposed by Donfack-Siewe et al.~\cite{DonfackSiewe2026_2}. Specifically, the ABC-$k$NN algorithm is implemented, by using $N$ numerical random field simulations to explore the prior space and selecting the $k$ nearest neighbors to approximate the multi-variable posterior density $\pi(\boldsymbol{\beta} \mid \widetilde{Z}_{\mathrm{obs}})$.

    While the baseline work by \cite{DonfackSiewe2026_2} details the application of this framework to Mode I fracture (DCB)—characterized by the hyperparameter vector $\boldsymbol{\beta}_I =(\mu_{G_{Ic}}, \mathrm{S}_{G_{Ic}}, \mu_{\sigma_{n,c}}, \mathrm{S}_{\sigma_{n,c}}, \ell_c^I)$ as shown in Figure \ref{result:fig1}—the present study extends this methodology to Mode II, utilizing the \textbf{End-Notched Flexure (ENF)} configuration. The Mode II parameters are defined by the vector $\boldsymbol{\beta}_{II}=( \mu_{G_{IIc}}, \mathrm{S}_{G_{IIc}}, \mu_{\sigma_{t,c}}, \mathrm{S}_{\sigma_{t,c}}, \ell_c^{II})$, with the corresponding identification results presented in Figure \ref{result:fig2}. 

    An analysis of the resulting posterior distributions for Mode II reveals that while certain hyperparameters shift from their uniform priors, others display less pronounced updates. This behavior is primarily attributed to the heightened non-linearity of the Mode II shearing mechanisms compared to Mode I, coupled with a more restricted computational budget ($N = 2997$) allocated for the ENF configuration due to convergence penalties. Even within the limitations of a reduced sample size, the likelihood-free framework identifies an exploitable spatial correlation scale.
    
    Consequently, the comprehensive set of interface parameters for the multi-scale stochastic modeling is defined as:
\begin{equation}
\boldsymbol{\beta} = (\mu_{G_{Ic}}, \mathrm{S}_{G_{Ic}}, \mu_{\sigma_{n,c}}, \mathrm{S}_{\sigma_{n,c}}, \ell_c^I, \mu_{G_{IIc}}, \mathrm{S}_{G_{IIc}}, \mu_{\sigma_{t,c}}, \mathrm{S}_{\sigma_{t,c}}, \ell_c^{II}).
\end{equation}
    
    To verify the physical representativeness of these calibrated parameters, Figure \ref{result:fig3} isolates the influence of spatial heterogeneity on the structural response. Unlike the global probabilistic confidence intervals, these results are generated using a fixed set of hyperparameters $\mu_{\boldsymbol{\beta}}  = \mathbb{E}_{\boldsymbol{\beta}  \sim \pi(\cdot \mid \widetilde{Z}_{\mathrm{obs}})}[\boldsymbol{\beta}]$, corresponding to the mean values of the posterior distribution. These mean values represent the identification targets to be propagated to the upper structural scale. A major advantage of the Bayesian framework is its ability to directly provide the explicit confidence bounds and estimation errors associated with these point estimates.

\begin{figure}[H]
    \centering
    \hspace*{-1.5cm}    
    \includegraphics[height=0.53\textheight]{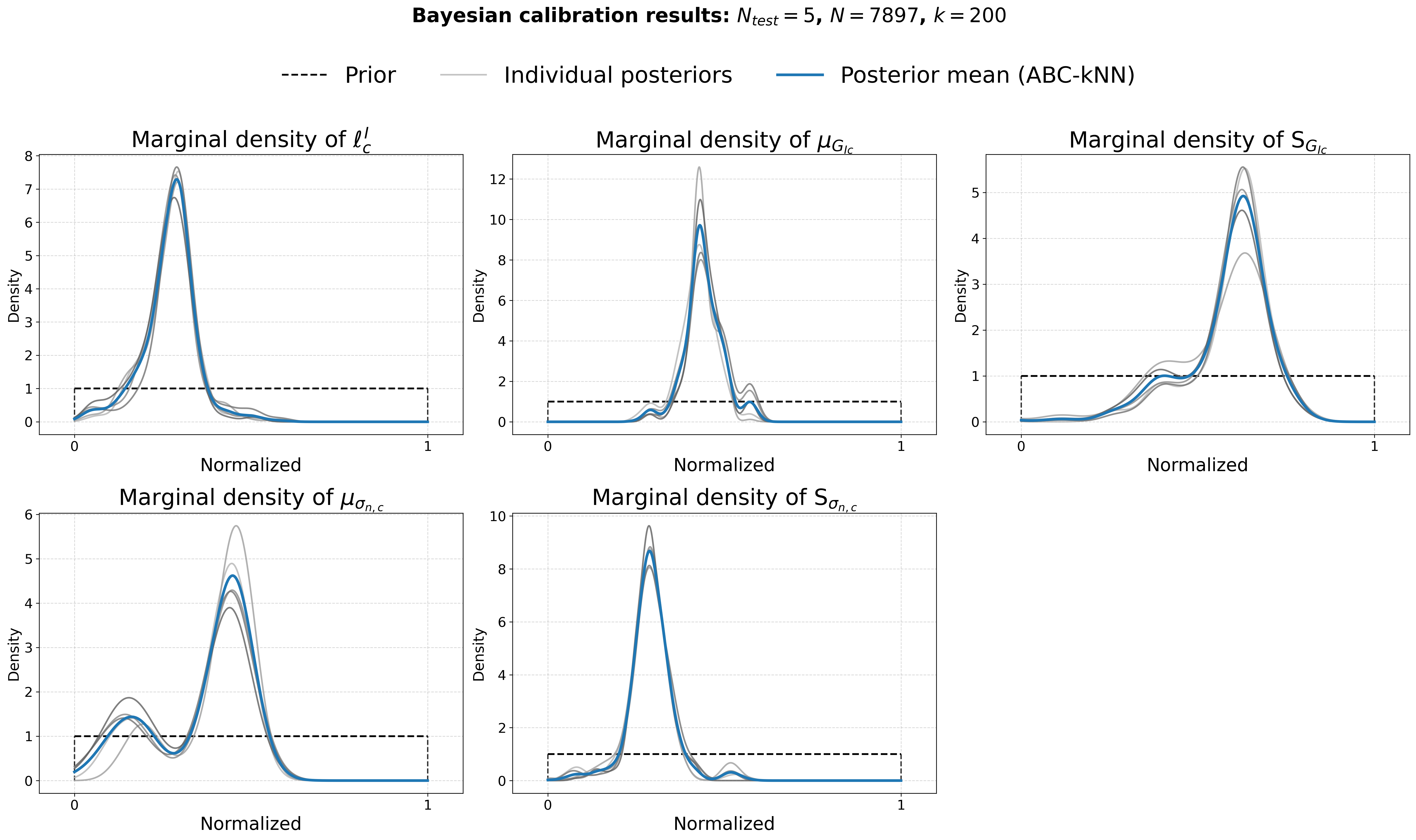}
    \caption{Posterior distributions of Mode I random field parameters calibrated from $N_{\mathrm{test}}=5$ DCB tests. The inference is based on a computational budget of $N=7897$ simulations and $k=200$ nearest neighbors \cite{DonfackSiewe2026_2}.}
    \label{result:fig1}
\end{figure}

\begin{figure}[H]
    \centering
    \hspace*{-1.5cm}    \includegraphics[height=0.8\textheight]{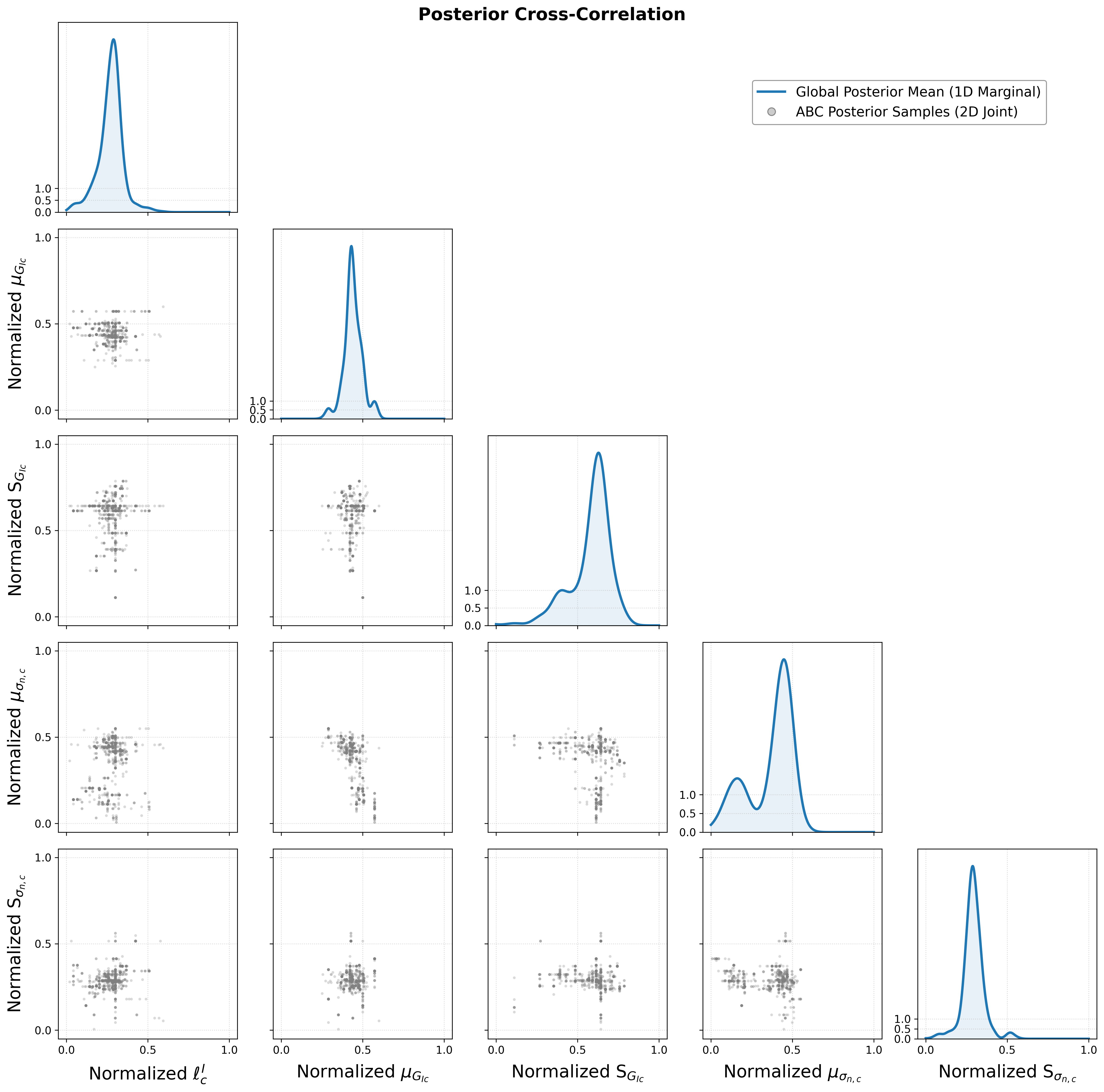}
    \caption{Posterior cross-correlation matrix for Mode I random field hyperparameters. The diagonal plots show the global 1D marginal posterior distributions, while the lower triangle scatter plots illustrate the pairwise 2D joint posterior distributions of the accepted ABC samples \cite{DonfackSiewe2026_2}.}
    \label{fig:cross_correlation}
\end{figure}

\begin{figure}[H]
    \centering
    \hspace*{-1.5cm}    
    \includegraphics[height=0.53\textheight]{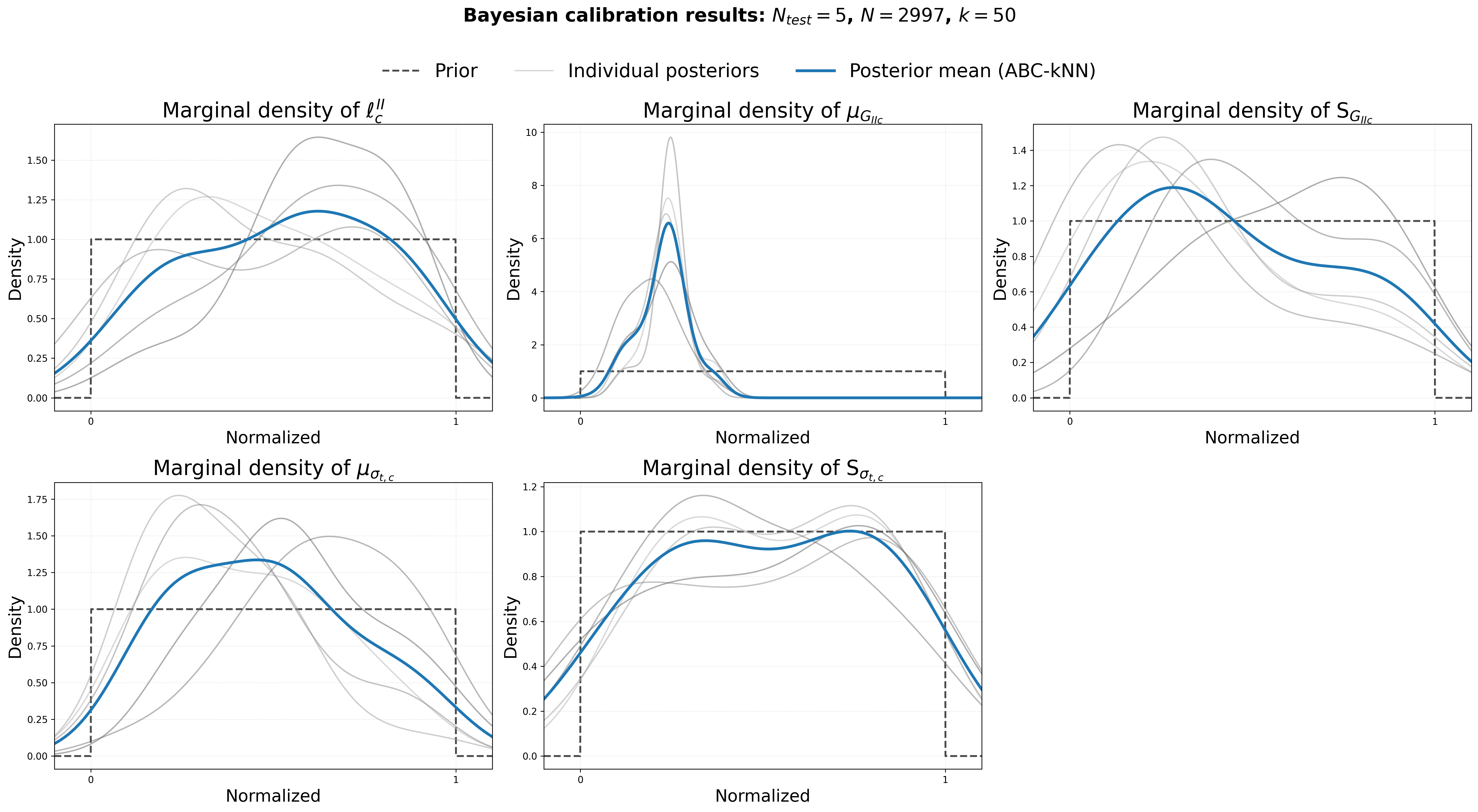}
    \caption{Posterior distributions of Mode II random field parameters calibrated from $N_{\mathrm{test}}=5$ ENF tests, using a total of $N=2997$ simulations and $k=50$ nearest neighbors.}
    \label{result:fig2}
\end{figure}

\begin{figure}[H]
    \centering
    \begin{subfigure}[b]{0.48\textwidth}
    \centering
    \includegraphics[width=\textwidth]{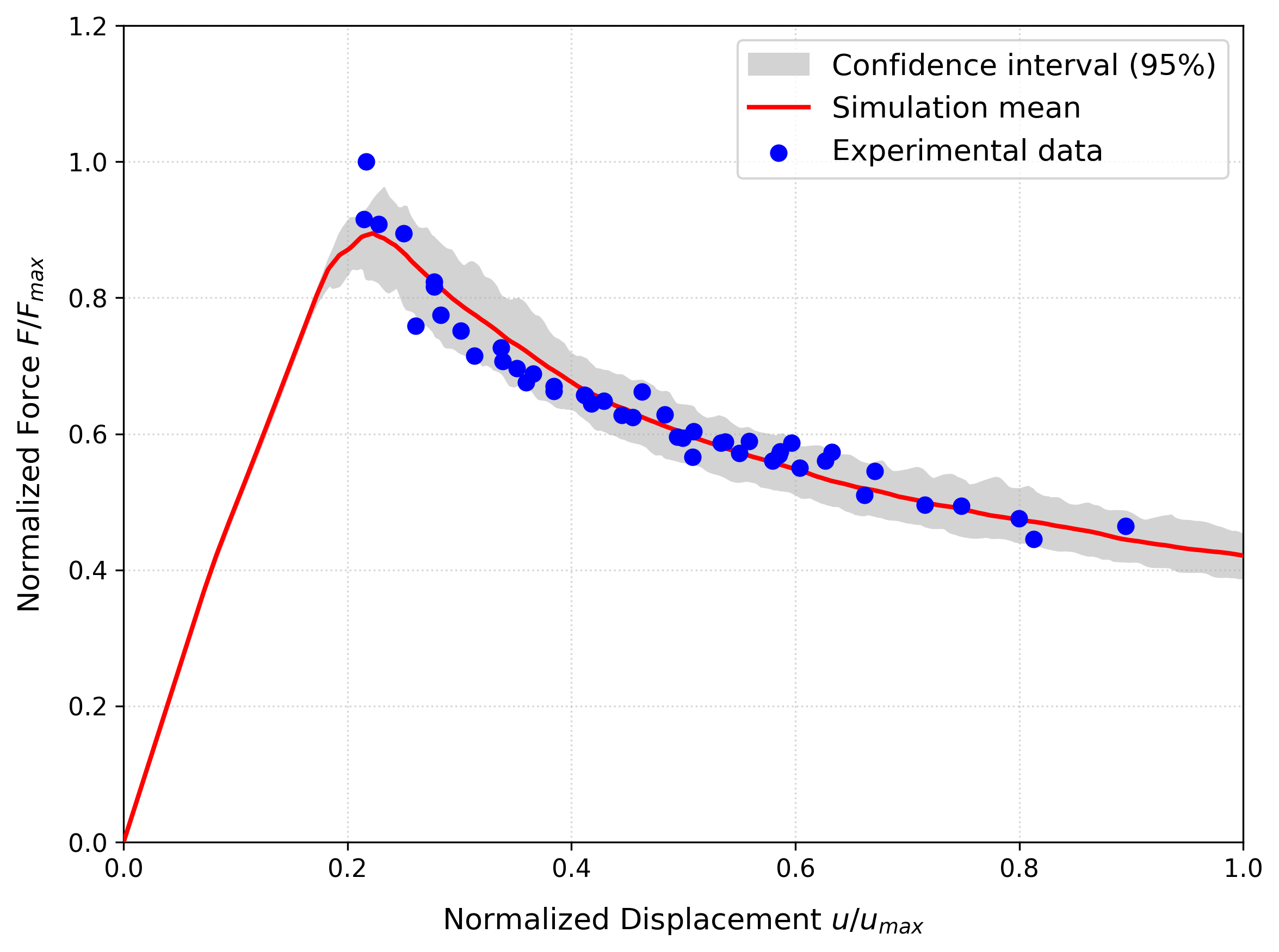}
    \caption{Mode I (DCB)}\label{fig:dcb_val}
    \end{subfigure}
    \hfill
    \begin{subfigure}[b]{0.48\textwidth}
    \centering
    \includegraphics[width=\textwidth]{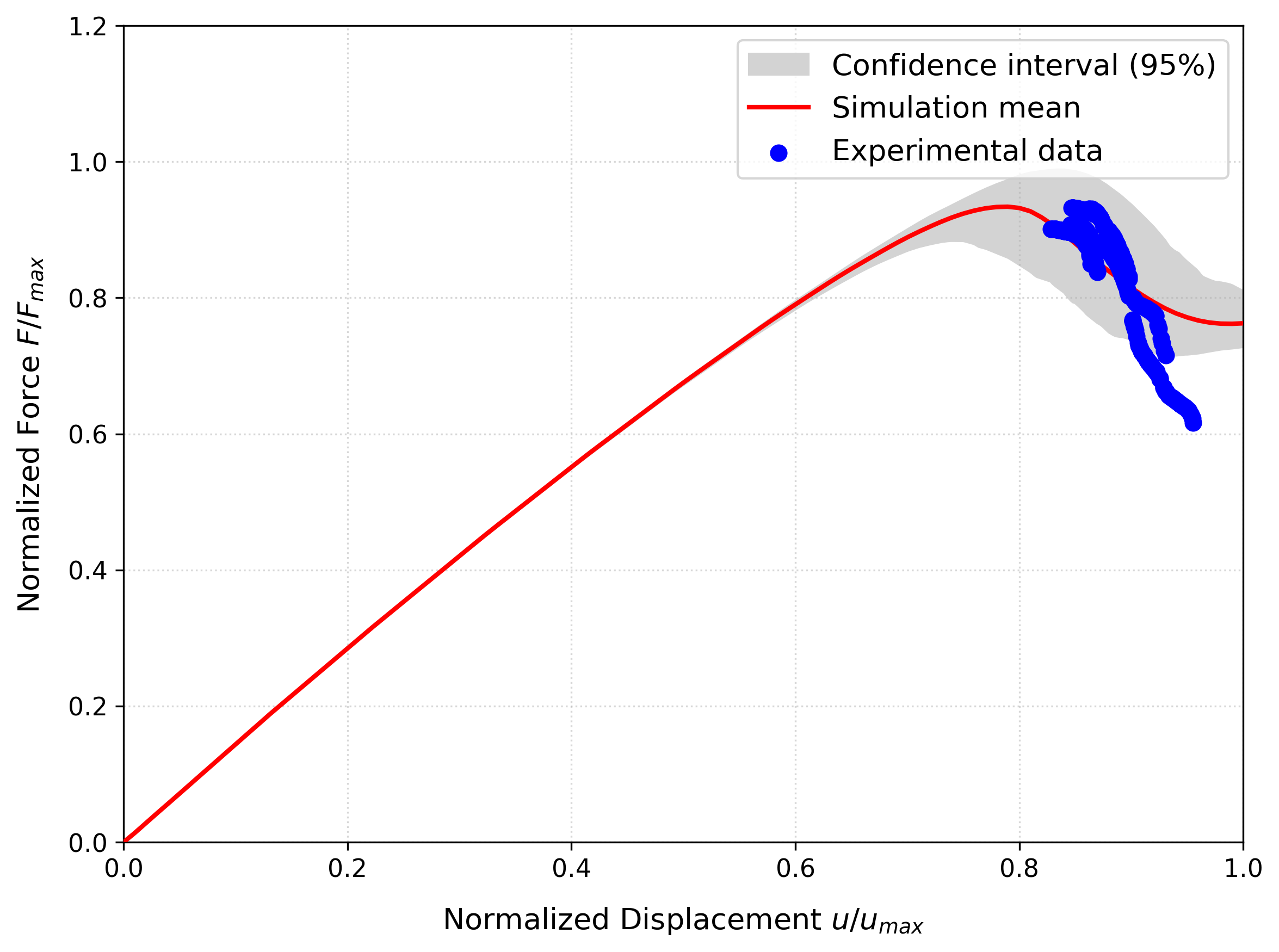}
    \caption{Mode II (ENF)}\label{fig:enf_val}
    \end{subfigure}
    \caption{Forward propagation of intrinsic variability: 100 numerical realizations (grey lines) generated from the posterior distribution mean $\mu_\Theta$, compared against experimental data (blue dots).}\label{result:fig3}
\end{figure}

\end{enumerate}

In summary, a comparative assessment between the space-invariant formulation (Figure \ref{Val_MCMC}) and the random field approach (Figure \ref{result:fig3}) visually shows that the integration of spatial random fields yields a significantly superior coverage and envelope of the experimental fracture data. Building upon these calibrated coupon-scale properties, the following section investigates scale-up effects at the upper structural level by systematically contrasting the predictive capabilities of these three uncertainty management paradigms.

\section{Computation of structural response variability at macroscale}
\label{section_4}
The transition from coupon-level identification to industrial-scale structural assessment necessitates an upscaling framework. This section details the application of the identified stochastic parameters to a stiffened aerospace panel (described in Section \ref{section_2}) investigating how interface variabilities influence the global structural response for this use case. Table \ref{tab:stochastic_params} summarizes the input stochastic parameters and the resulting structural outputs. The  coefficients of variation ($\text{CoV}$) are evaluated using the non-normalized data, thereby providing a direct representation of the real material variability inherent to each scale.

\begin{table}[H]
    \caption{Summary of stochastic inputs and structural response outputs obtained from 50 Monte Carlo simulations (MCs). All structural outputs are normalized by the maximum force $F_{max}$ of all the simulations. Values in square brackets $[...]$ denote the variability range observed across realizations. The parameter $s$ represents the characteristic mesh size, and CoV denotes the Coefficient of Variation.}
    \label{tab:stochastic_params}
    \noindent\hspace*{-1.5cm}
    \begin{tabular}{|l|c|c|}
        \hline
        \rowcolor{gray!20} \textbf{Parameter} & \textbf{Paradigm 2} & \textbf{Paradigm 3} \\
        \hline  \multicolumn{3}{|l|}
        {\textit{Input Stochastic Parameters  }}  \\
        \hline
        $G_{Ic}$ (CoV) &  $2.46\%$ & $25.27\%$   \\
        \hline
        $G_{IIc}$ (CoV) &  $23.86\%$ & $18.96\%$   \\
        \hline
        $\sigma_{n,c}$ (CoV) &  $16.37\%$ & $15.48\%$   \\
        \hline
        $\sigma_{t,c}$ (CoV) &  $9.57\%$ & $3.90\%$   \\
        \hline
        $l_{c}^{I}/s$ & N/A & $0.6$   \\
        \hline
        $l_{c}^{II}/s$ & N/A & $1.21$   \\
        \hline
        \hline
        \rowcolor{blue!10} \multicolumn{3}{|l|}{\textit{Structural Response (Outputs)  }} \\
        \hline
        \rowcolor{blue!5} Skin buckling force & $0.30$ (Constant) & $0.30$ (Constant) \\
        \hline
        \rowcolor{blue!5} Initial disbondingforce &   $0.5\%$, $\text{Range}= [0.93 - 0.96]$   &   $10.9\%$, $\text{Range}= [0.63 - 0.94]$\\
        \hline
        \rowcolor{blue!5} Skin buckling force under stringer &   $1.4\%$, $\text{Range}= [0.88 - 0.95]$   &   $0.4\%$, $\text{Range}= [0.93 - 0.94]$\\
        \hline
        \rowcolor{blue!5} Global peak force &   $1.1\%$, $\text{Range}= [0.95 - 1.00]$  &   $0.21\%$, $\text{Range}= [0.97 - 0.98] $ \\
        \hline
    \end{tabular}
\end{table}

\subsection{Results and discussion: scale effects and attenuation}

The structural response curves illustrated in Figure~\ref{comparison1} and the corresponding spatial footprints of disbondinginitiation in Figure~\ref{fig:spatial_maps_comparison} outline the progression of structural collapse. As a reference baseline, the traditional deterministic approach using a B-basis value demonstrates conservatism, underpredicting both the damage initiation threshold and the ultimate load capacity. By contrast, the probabilistic paradigms provide a comprehensive timeline of failure kinematics, unfolding sequentially through distinct mechanical transitions governed by the interplay between global geometry and local interface properties.

For this use case, the failure sequence begins with local skin buckling between the two stringers, initiating early at a uniform normalized load of $0.30$ across all stochastic realizations in Table \ref{tab:stochastic_params}. This early buckling is not impacted by the interface variability, yielding a coefficient of variation ($\text{CoV}$) of $0.0\%$. 

As compressive displacement increases, the post-buckled skin redistributes the mechanical loads, transmitting severe shear and peeling stresses directly toward the skin-stringer bond lines. This triggers interfacial damage initiation, defined by the localized variable reaching unity ($d_{max} = 1$). For this initiation event, Table \ref{tab:stochastic_params} highlights a fundamental divergence between the uncertainty paradigms. The constant-property baseline assumes perfect spatial homogeneity from panel to panel, resulting in a delayed, nearly deterministic damage onset ($[0.93 - 0.96]$, $\text{CoV} = 0.5\%$). Conversely, the proposed random field paradigm captures premature local degradation driven by spatial weak spots, showing a large initiation dispersion ($[0.63 - 0.94]$, $\text{CoV} = 10.9\%$).

The structural implications of these distinct damage initiation paths are closely linked to their spatial localization, which is best understood by contrasting the two probabilistic approaches:
\begin{itemize}
    \item \textbf{Paradigm 2:} As illustrated in the Figure \ref{fig:spatial_map_sc_comp}, this approach reveals a highly localized and uniform initiation behavior, completely devoid of spatial dispersion. This means that initiation occurs simultaneously in the most critical high-gradient areas, located at the inner radius of the stringers. Consequently, this configuration bypasses any progressive accumulation of local damage and leads directly to an ideal and extended propagation path.
    \item \textbf{Paradigm 3:} As illustrated in Figure~\ref{fig:spatial_map_field_comp}, this approach exhibits a scattered damage initiation pattern across the spatial domain. In these random field realizations, disbondingis "captured" by local weak spots within the interface. Analysis reveals that initiations appearing in the skin between the stringers are relatively stable; they do not propagate significantly and have a minimal influence on global integrity. Conversely, initiations occurring between the skin and stringer in the inner radius of the stringers are critical. In these regions, the buckling of the skin is accompanied by significantly elevated interface stresses. The coincidence of these high-stress gradients with a local stochastic weak spot triggers rapid, unstable crack propagation and ultimate panel rupture. This propagation creates a fully debonded zone between the skin and the stringer. Ultimately, the use of random fields demonstrates that the critical driver for this panel integrity is the buckling of the skin under the stringer coupled with disbondinginitiation, whereas the intrinsic variability of interface properties has a negligible effect on the global peak load.
\end{itemize}
Despite these distinct localized damage initiation paths, both paradigms ultimately converge toward a stable global peak force for this design characterized by low coefficient of variation. Specifically, the ultimate failure load exhibits a $\text{CoV}$ of only $0.21\%$ for the random field paradigm and $1.1\%$ for the constant-property baseline. 

\subsection{Uncertainty attenuation and industrial takeaways for this specific design}

\begin{figure}[H]
    \hspace{-2.6cm}
    \includegraphics[width=1.45\textwidth]{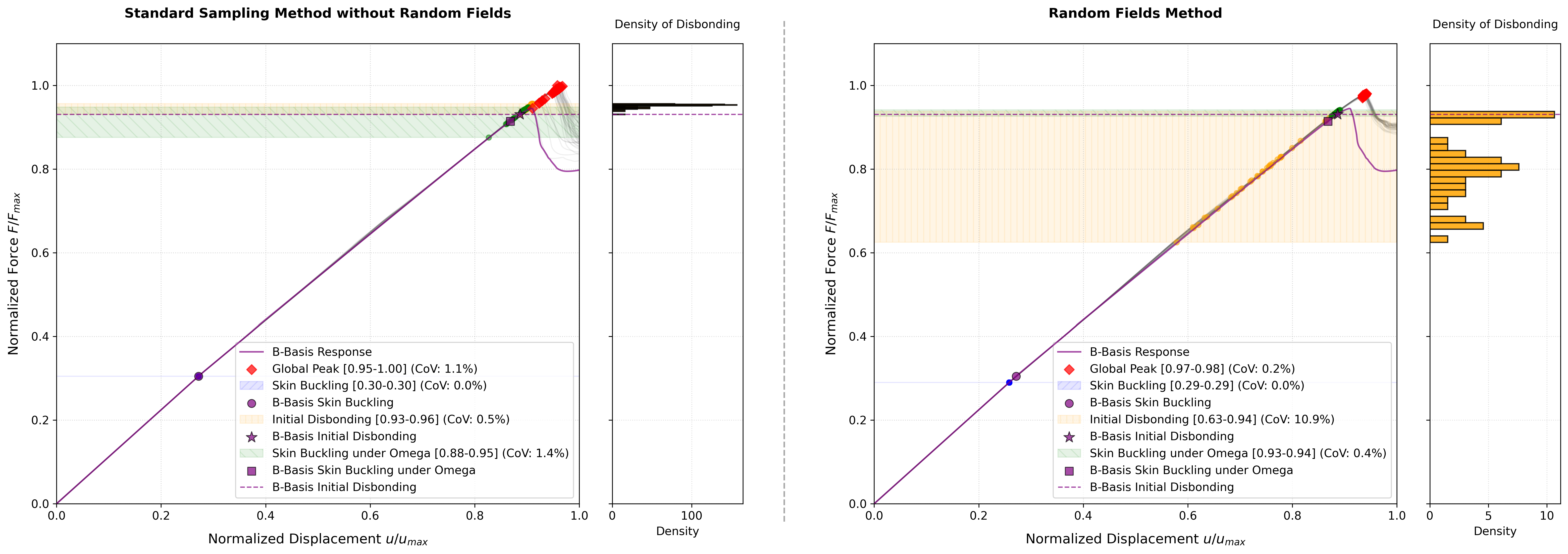}
    \caption{Stochastic global (50 MCS) response and failure sequence comparison between the standard sampling method (left) and the random fields method (right). The plots display the normalized force-displacement curves, the variability ranges with coefficients of variation (CoV) for key instability events, the purple B-Basis reference limits, and the corresponding probability density histograms for the initial disbonding force.}
    \label{comparison1}
\end{figure}

\begin{figure}[H]
    \centering
    \begin{subfigure}[b]{0.48\textwidth}
        \centering
        \includegraphics[width=\textwidth]{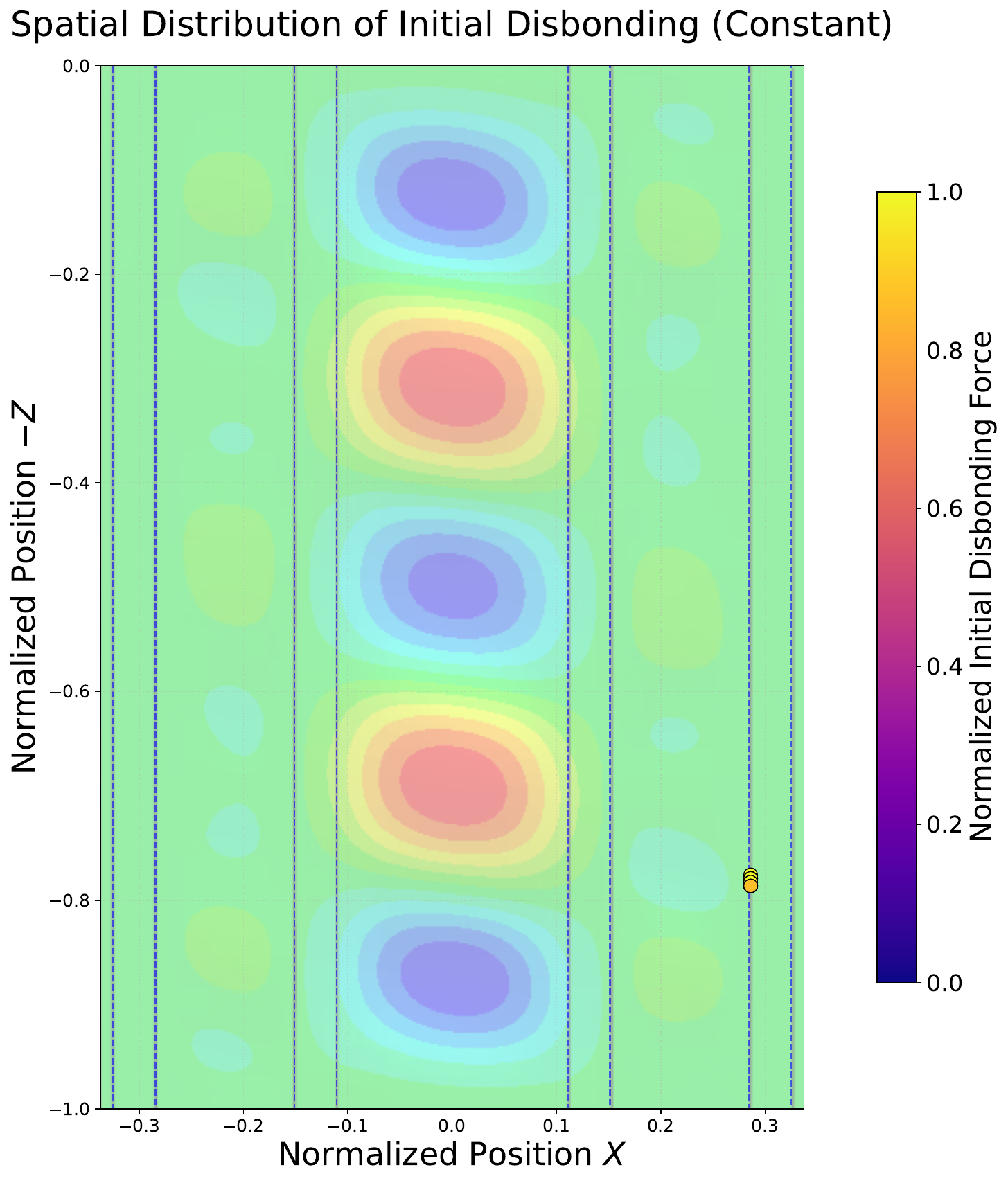}
        \caption{Baseline method}
        \label{fig:spatial_map_sc_comp}
    \end{subfigure}
    \hfill
    \begin{subfigure}[b]{0.48\textwidth}
        \centering
        \includegraphics[width=\textwidth]{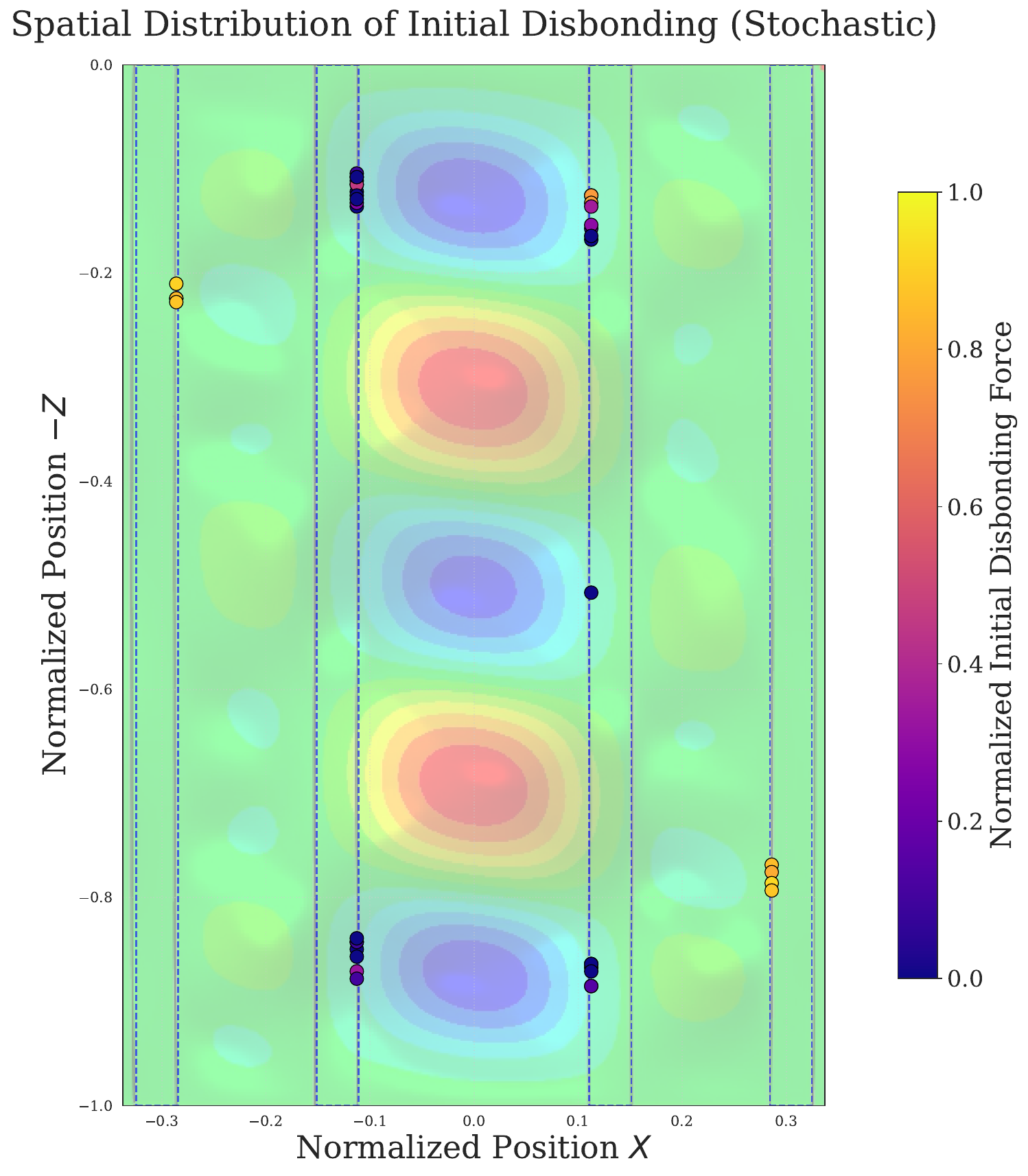}
        \caption{Random fields method}
        \label{fig:spatial_map_field_comp}
    \end{subfigure}
    \vspace{10pt}
    \caption{Comparison of the spatial distribution of initial disbondinglocations (50 MCS): (a) constant approach versus (b) random fields approach. The locations are overlaid on the buckling mode shape, where the contour represents out-of-plane displacement (red and blue indicate positive and negative values, respectively).}
    \label{fig:spatial_maps_comparison}
\end{figure}

A fundamental finding of this numerical study for this design is the observed \textbf{scale effect} in uncertainty propagation (Figure \ref{comparison1}). Despite the significant scatter in interfacial fracture properties identified at the coupon scale (Table \ref{tab:stochastic_params}), the global structural response of the stiffened panel remains remarkably stable when spatial variability is considered. This phenomenon suggests that the structural scale provides a form of \textbf{statistical averaging} or attenuation, where local heterogeneities primarily govern the specific spatial location of disbondinginitiation but do not significantly alter the global panel stiffness or ultimate load-bearing capacity. This type of analysis helps engineers determine whether a higher level of material variability can be safely tolerated at the coupon scale, provided that the upper structural level displays a total averaging phenomenon—as demonstrated by the remarkably stable $0.21\%$ $\text{CoV}$) for the ultimate load compared to the $10.9\%$ $\text{CoV}$ at initiation. Consequently, conventional design criteria and safety margins could be optimized, avoiding excessive weight penalties for global failure responses that remain nearly deterministic despite local material fluctuations.

Another significant advantage of this multi-scale numerical approach arises when the sizing criterion or critical design driver for the structure is stringer debonding. In such scenarios, modeling via spatially correlated random fields becomes indispensable, as it captures potential damage initiation zones that are completely overlooked by classical uniform formulations. This predictive capability of modeling via spatially correlated random fields (Paradigm 3) offers advantages for Structural Health Monitoring (SHM) \cite{giurgiutiu2014structural} strategies in damage-tolerant airframes. While global mechanics dictate the ultimate panel collapse, Paradigm 3 isolates and maps spatial topology of early damage initiation zones. From an aircraft engineering perspective, this allows for the transition from empirical sensor placement to a physics-based, digital-twin-driven deployment strategy. Knowing precisely where micro-disbonding is statistically bound to originate enables engineers to optimize the layout of localized sensor networks (such as piezoelectric patches or Fiber Bragg Grating optical fibers) prior to ground testing. This strategy aligns with advanced sensor placement procedures that incorporate the spatial correlation length of material defects to maximize detection capability at minimum cost \cite{ClercSchoefs2019}.
For this specific panel design and configuration, early damage initiation alone does not constitute the critical mechanism leading to structural collapse ; instead, final failure is strictly governed by global mechanics—specifically, the severe out-of-plane peeling stresses induced by skin buckling beneath the stringers.  Other configurations or structural designs might exhibit a heightened sensitivity to early debonding. Ultimately, by moving beyond idealized, homogeneous failure modes, this methodology establishes a pathway toward test derisking through high-fidelity digital twins. These predictive tools can numerically identify where damage initiation is most probable, offering a strategic advantage for optimized sensor placement and structural health monitoring during experimental testing campaigns.

\section{Conclusion}
This study established a multi-scale stochastic framework to numerically investigate the impact of interface property variability in composite structures on the structural response at a larger scale using three paradigms. While the developed computational methodology offers a generic, universally applicable toolchain for any composite assembly, the resulting structural mechanics are highly specific to the investigated configuration. The transition from coupon-level identification to industrial-scale components highlights a clear dual message for aerospace design engineers.  

On one hand, regarding global design criteria (such as the skin buckling force and the ultimate force), local failure property fluctuations due to manufacturing are filtered out by a scale-driven statistical averaging effect. Consequently, the global load capacity remains remarkably stable, meaning that a simpler, computationally lightweight uniform stochastic approach (Paradigm 2) provides a sufficient approach for macroscopic sizing.

On the other hand, for local design criteria (such as stringer disbonding initiation and damage tracking), the assumption of spatial homogeneity fails to capture local reality. In this specific context, modeling via spatially correlated random fields (Paradigm 3) becomes indispensable, as it isolates early damage initiation zones triggered by the coincidence of high stress gradients with local stochastic weak spots.

This methodology paves the way for a transition from deterministic design approaches to physics-based, high-fidelity virtual certification and simulation-based uncertainty propagation. Within this virtual testing framework, the macroscale finite-element model directly incorporates the cohesive laws and conforming interface discretization already calibrated the coupon level, providing an objective predictive platform for test derisking prior to physical qualification. Future work aims to integrate global manufacturing uncertainties into the random field formalism to achieve a fully integrated digital twin of composite airframes.

Further research could also explore the estimation of panel-scale numerical design values—such as quantiles of the skin buckling force, stringer disbonding initiation, and ultimate global buckling load—by leveraging limited experimental data from both coupon and structural levels~\cite{DonfackSiewe2026_1}. This involves accounting for epistemic uncertainties arising from the limited availability of experimental data used to identify input distributions. Finally, addressing the computational challenges related to the high number of simulations required for rigorous uncertainty propagation will be essential to ensure the robustness and practical applicability of these frameworks.

\section*{Declaration of competing interest}
The authors declare that they have no known competing financial interests or personal relationships that could have appeared to influence the work reported in this paper.

\section*{Data Availability Statement}
The normalized experimental test datasets (Mode I DCB and Mode II ENF force--displacement curves), along with the Python source code and post-processing scripts supporting the calibration and validation results of this study, are openly available in the Zenodo repository at \url{https://doi.org/10.5281/zenodo.22199632}

\appendix

\section{Experimental dataset and test results}
\label{app:experiments}

The experimental dataset provided by Airbus, comprising standardized Mode I (DCB) and Mode II (ENF) fracture tests. The campaign involves five independent specimens for each configuration ($N_{\text{test}}=5$).

The specimens are manufactured from carbon-epoxy unidirectional (UD) laminates with a stacking sequence of $[0^\circ]_8 // [0^\circ]_8$. Each arm consists of 8 plies with a cured ply thickness (CPT) of 0.184 mm, resulting in a total specimen thickness of approximately 3 mm (see Table \ref{tab:layup}).

\begin{figure}[H]
    \centering
    \includegraphics[width=0.7\textwidth]{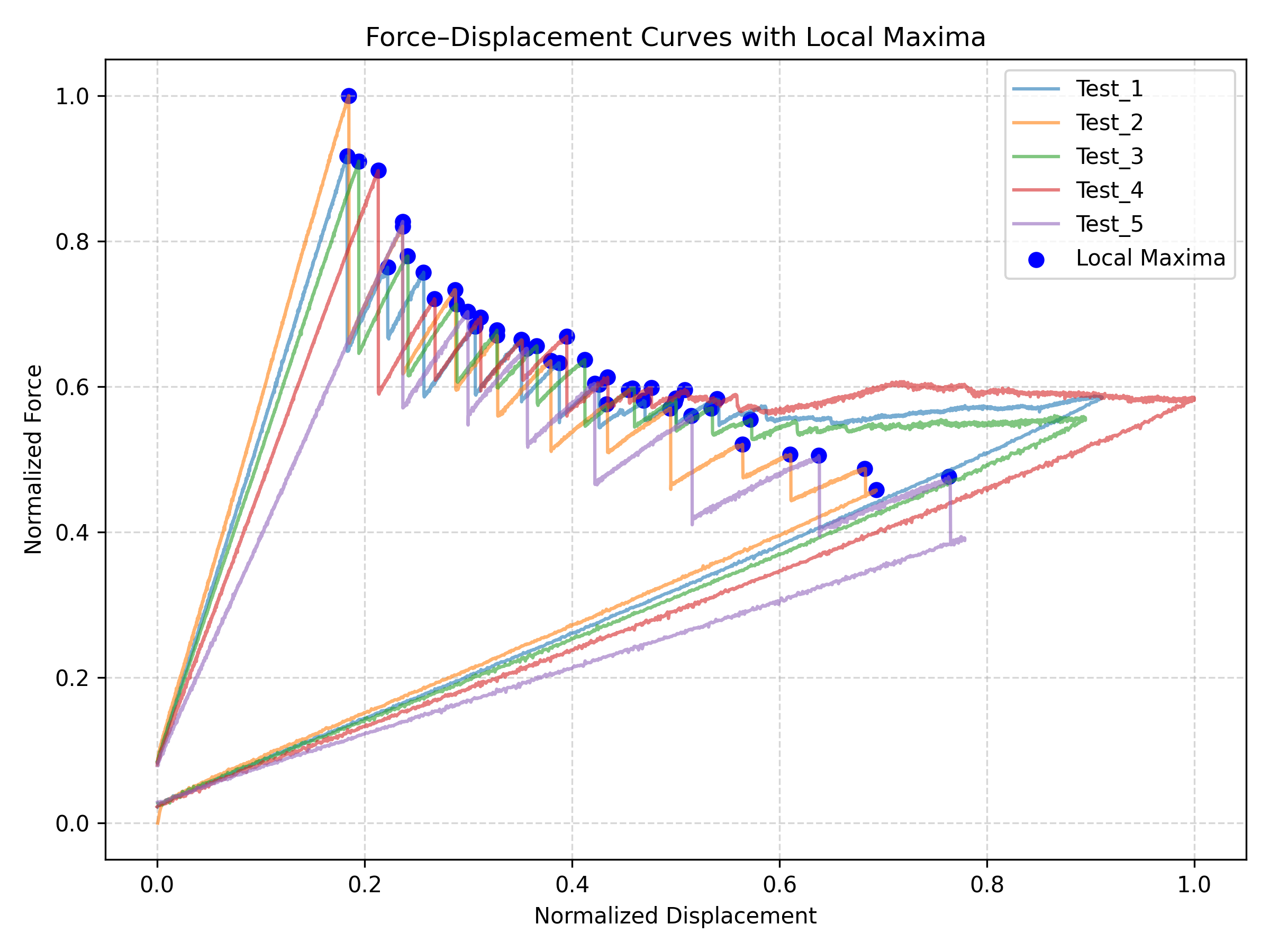}
    \caption{Normalized experimental response of the five DCB specimens (Mode I). Blue dots indicate local maxima associated with stick-slip crack propagation.}
    \label{fig:dcb_exp}
\end{figure}

\begin{figure}[H]
    \centering
    \includegraphics[width=0.7\textwidth]{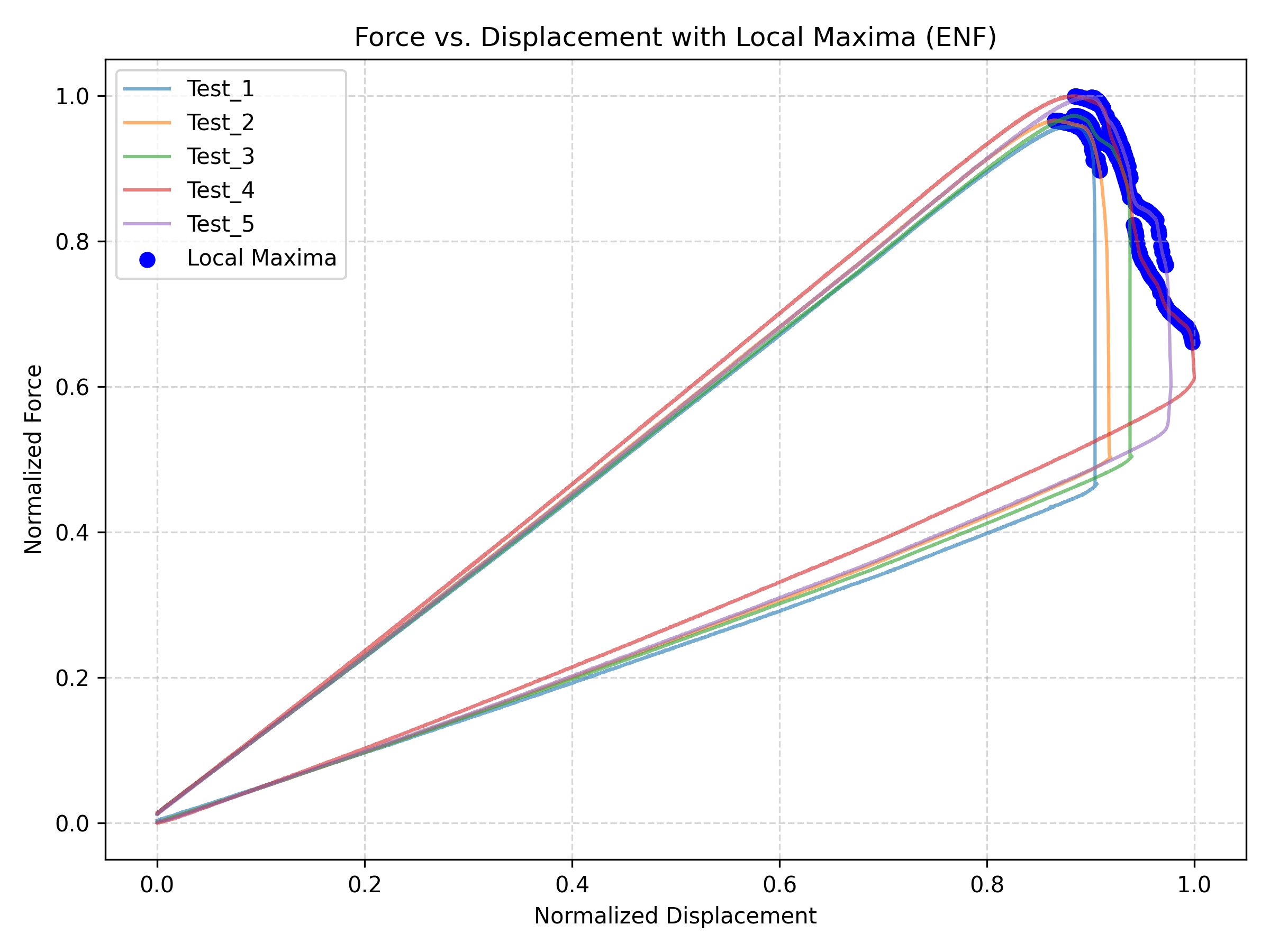}
    \caption{Normalized experimental response of the five ENF specimens (Mode II). The dispersion highlights the influence of interface heterogeneity on the macroscopic response.}
    \label{fig:enf_exp}
\end{figure}

Figures \ref{fig:dcb_exp} and \ref{fig:enf_exp} present the normalized experimental responses. 
\begin{table}[h]
    \centering
    \begin{tabular}{|c|c|c|c|}
        \hline
        \textbf{Stacking sequence} & \textbf{CPT [mm]} & \textbf{Number of plies} & \textbf{Total thickness [mm]} \\
        \hline
        $[0^\circ]_8 ~\|~ [0^\circ]_8$ & 0.184 & 16 & 2.994 \\
        \hline
    \end{tabular}
    \caption{Laminate configuration used for all DCB and ENF specimens.}
    \label{tab:layup}
\end{table}


\bibliographystyle{elsarticle-num} 
\bibliography{Mabib}





\end{document}